\documentclass[%
twocolumn,
amsmath,amssymb,
aps,
superscriptaddress,
]{revtex4-2}

\usepackage{url}
\usepackage{graphicx}
\graphicspath{{.}{figs/}{../figs/}}
\usepackage{afterpage}
\usepackage{siunitx}
\DeclareSIUnit\angstrom{\text {Å}}
\usepackage{dcolumn}
\usepackage{bm}
\usepackage{tikz}
\usetikzlibrary{positioning,intersections,angles,quotes}
\usepackage[caption=false]{subfig}
\usepackage{booktabs} 
\usepackage{comment}
\usepackage{physics}
\usepackage{mathtools}

\usepackage[colorlinks=true,citecolor=blue,linkcolor=magenta]{hyperref}

\usepackage{longtable}
\usepackage{booktabs}
\usepackage{svg}
\usepackage{xcolor} 
\begin{document}

\title{VQE-generated quantum circuit dataset for machine learning}

\author{Akimoto Nakayama}
\affiliation{%
  Graduate School of Engineering Science, Osaka University, 1-3 Machikaneyama, Toyonaka, Osaka 560-8531, Japan
}
\author{Kosuke Mitarai}
\email{mitarai.kosuke.es@osaka-u.ac.jp}
\affiliation{%
  Graduate School of Engineering Science, Osaka University, 1-3 Machikaneyama, Toyonaka, Osaka 560-8531, Japan
}%
\affiliation{%
  Center for Quantum Information and Quantum Biology,
  Osaka University, 1-2 Machikaneyama, Toyonaka 560-0043, Japan
}%
\author{Leonardo Placidi}
\affiliation{%
  Graduate School of Engineering Science, Osaka University, 1-3 Machikaneyama, Toyonaka, Osaka 560-8531, Japan
}
\affiliation{%
  Center for Quantum Information and Quantum Biology,
  Osaka University, 1-2 Machikaneyama, Toyonaka 560-0043, Japan
}%
\author{Takanori Sugimoto}
\affiliation{%
  Center for Quantum Information and Quantum Biology,
  Osaka University, 1-2 Machikaneyama, Toyonaka 560-0043, Japan
}%
\affiliation{
Computational Materials Science Research Team, RIKEN Center for Computational Science (R-CCS), Kobe, Hyogo, 650-0047, Japan
}
\affiliation{
Advanced Science Research Center, Japan Atomic Energy Agency, Tokai, Ibaraki 319-1195, Japan
}

\author{Keisuke Fujii}%
\email{fujii@qc.ee.es.osaka-u.ac.jp}
\affiliation{%
   Graduate School of Engineering Science, Osaka University, 1-3 Machikaneyama, Toyonaka, Osaka 560-8531, Japan
}%
\affiliation{%
  Center for Quantum Information and Quantum Biology,
  Osaka University, 1-2 Machikaneyama, Toyonaka 560-0043, Japan
}%
\affiliation{%
  RIKEN Center for Quantum Computing (RQC),
  Hirosawa 2-1, Wako, Saitama 351-0198, Japan
}%
%
\date{\today}
\begin{abstract}
Quantum machine learning has the potential to computationally outperform classical machine learning, but it is not yet clear whether it will actually be valuable for practical problems. While some artificial scenarios have shown that certain quantum machine learning techniques may be advantageous compared to their classical counterpart, evidence does not yet suggest that quantum machine learning has surpassed conventional approaches in dealing with standard classical datasets, such as MNIST dataset. In contrast, dealing with quantum data, such as quantum states or circuits, may be the task where we can benefit from quantum methods. Therefore, it is important to develop practically meaningful quantum datasets for which we expect quantum methods to be superior. In this paper, we propose a machine learning task that is likely to soon arise in the real world: clustering and classification of quantum circuits. We provide a dataset of quantum circuits optimized by the variational quantum eigensolver. 
We utilized six common types of Hamiltonians in condensed matter physics, with a range of 4-20 qubits, and applied ten different ansatz with varying depths (ranging from 3 to 32) to generate a quantum circuit dataset of six distinct classes, each containing 300 samples.
We show that this dataset can be easily learned using quantum methods. In particular, we demonstrate a successful classification of our dataset using real four-qubit devices available through IBMQ. By providing a setting and an elementary dataset where quantum machine learning is expected to be beneficial, we hope to encourage and ease the advancement of the field.
\end{abstract}

\maketitle

\section{Introduction}

Quantum machine learning has attracted much attention in recent years as a promising application of quantum computers \cite{biamonte2017quantum, cerezo2022challenges}.
Many techniques, such as quantum neural networks \cite{farhi2018, mitarai2018quantum, Schuld2019}, quantum generative models \cite{Liu2018, Benedetti2019}, quantum kernel methods \cite{havlivcek2019supervised}, and so on have been developed for achieving possible quantum speedups in machine learning tasks.
They have also been realized experimentally \cite{Benedetti2019, havlivcek2019supervised, Kusumoto2021, bartkiewicz2020experimental, PhysRevX.12.031010}.
While some artificial, carefully-designed scenarios have demonstrated that certain quantum machine learning techniques may be advantageous compared to classical methods \cite{Liu_2021, Huang2021, dunjko2017, NEURIPS2021_eec96a7f, Pirnay2022}, it is not yet clear whether quantum techniques would be beneficial for practical applications.

In traditional machine learning, standard datasets, such as MNIST handwritten digits \cite{Lecun1998}, are used to evaluate the performance and thus the practicality of new models. 
As of now, there is no evidence to suggest that quantum machine learning methods have surpassed the state-of-the-art classical machine learning techniques on these datasets. On the other hand, it has been demonstrated that classical machine learning can efficiently learn from these datasets. For example, with a large-scale numerical experiment involving up to 30 qubits, 
Huang \textit{et al.} \cite{Huang2021} have shown that the Fashion-MNIST dataset \cite{fashionmnist} is better learned by classical models.
In contrast, when working with “quantum data”, such as quantum states or circuits, there is a good reason to believe that quantum computers may provide a significant advantage. 
In another work by Huang \textit{et al.} \cite{Huang2022}, it has been rigorously shown that quantum machine learning is beneficial when learning unknown quantum states or processes provided from physical experiments.
It is therefore important to develop a practical quantum dataset in which we can expect quantum methods to be superior.

Several works have made efforts in this direction.
Schatski \textit{et al.} \cite{ntangled} proposed a dataset consisting of parameterized quantum circuits with different structure whose parameters are optimized to give output states with certain values of entanglement.
However, this dataset lacks data involving higher qubit counts, which is crucial for evaluating scalable quantum machine learning methods.
Also, Huang \textit{et al.} \cite{Huang2021} proposed to relabel a classical dataset by outputs of quantum circuits so that the relabeled one is difficult to be learned by classical computers.
These examples, while giving datasets that are possibly hard to learn by classical computers, are not plausible “real-world” quantum data.
Consequently, it remains unclear how insights or results from these datasets could translate into realistic scenarios or tangible real-world benefits.
Perrier, Youssry and Ferrie \cite{Perrier2022} provides a quantum dataset called QDataSet, but its aim is to provide a benchmark for classical machine learning applied to quantum physics, and therefore it does not fit to the context of this paper.  
An example of a quantum-classical dataset in which quantum and classical methods have relatively good performance is instead MNISQ~\cite{placidi2023mnisq}, which was introduced after the present work.
In our work, we apply machine learning techniques, such as classification or clustering, directly to quantum circuits as input data. 
Moreover, these quantum circuits originate from practically important quantum computations and are of sufficient scale to meaningfully evaluate their usefulness on current or near-future quantum hardware.

\begin{figure}
 \centering
 \includegraphics[width=\linewidth]{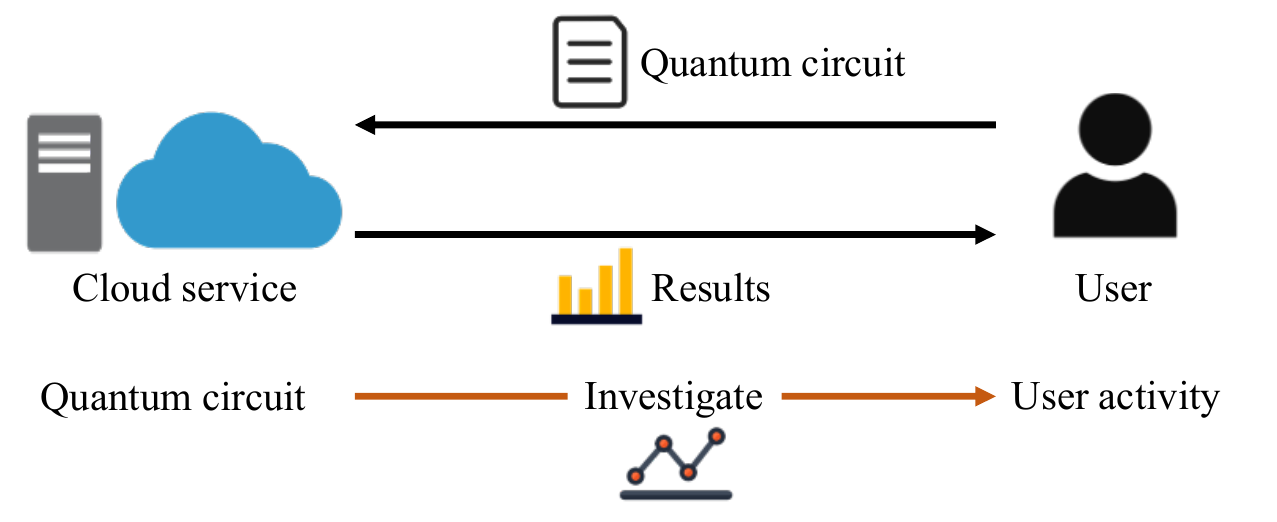}
 \caption{\textbf{Scenario of the proposed machine learning task.} Cloud quantum computer providers get descriptions of quantum circuits constantly from their users and return execution results to them. The providers wish to investigate the user activities, e.g., whether two users are interested in solving a similar computational task or not, from the circuit data.}
 \label{overview}
\end{figure}

In this work, we propose a more practical machine learning task that we expect to naturally arise in near the future and in the real-world: a clustering and classification of many quantum circuits, and provide an elementary dataset for this task.
A successful model that can perform such a task could be beneficial to providers of cloud quantum computers as depicted in Fig.~\ref{overview}, it would allow them to understand the preferences of their users by analyzing circuits submitted by them.
While there are many possible ways to analyze the circuits, in this work, we focus on a setting where the providers want to cluster or classify circuits based on their similarity of output states.
This task is easy when we have access to quantum computers because similarity, which can be measured in terms of overlaps between output states, can be readily estimated on them.
Such estimation is likely to be hard for classical computers when a quantum circuit is large enough \cite{doi:10.1137/S0097539796300921}.

The quantum dataset provided in this work is a set of quantum circuits optimized by the variational quantum eigensolver (VQE) \cite{Peruzzo_2014, TILLY20221}, which is an algorithm to find circuits that output approximate ground states of quantum systems. 
More specifically, we use six model Hamiltonians that are famous in condensed matter physics, and optimize 300 different parametrized circuits with varying structure and depth to output each of their ground states.
The dataset includes circuits with up to 20 qubits, but it can be easily extended to larger numbers of qubits with access to quantum hardware. To demonstrate the potential of the dataset, we perform a proof-of-principle experiment using quantum circuit simulators and show that the circuits can be accurately clustered.
We also demonstrate successful classification of our four-qubit dataset using real quantum hardware.

The dataset is freely accessible on the GitHub repository \cite{nakayama2023github}, therefore the reader will be able to freely use it for research or benchmarking. For each number of qubit, data are stored in QASM \cite{10.1145/3505636} format and are publicly accessible.
By providing a natural setting and an elementary dataset where quantum machine learning is expected to be beneficial, we hope to support the thoughtful and grounded advancement of the field.

\section{Dataset construction}

\subsection{Idea overview}

The machine learning task that we consider in this work is clustering and classification of quantum circuits based on the similarity of their output states.
More specifically, the task is to classify $M$ quantum circuits $\{U_m\mid m=1,2,\cdots,M\}$ based on the fidelities of output states $|\bra{0}U_m^\dagger U_{m^\prime}\ket{0}|^2$.
We expect this task to naturally arise when quantum computer providers wish to analyze what their users do on their hardware.
Also, we believe this task to be hard in general when we have access only to classical computers, since estimation of $|\bra{0}U\ket{0}|^2$ to an accuracy of $\epsilon$ for a general polynomial-sized quantum circuit $U$ in polynomial time is clearly a BQP-complete task.

To construct an elementary dataset for this task, we use the VQE \cite{Peruzzo_2014, TILLY20221}.
It is a method to generate a quantum circuit which outputs an approximate ground state of a target Hamiltonian $H$.
This is usually done by using a parameterized quantum circuit, also referred to as an ansatz, $U(\bm{\theta})$ whose parameter $\bm{\theta}$ is optimized to minimize the energy expectation value $\expval{H(\bm{\theta})} := \bra{0}U^\dagger(\bm{\theta})HU(\bm{\theta})\ket{0}$.
The dataset is constructed by optimizing various ansatz to generate ground states of different Hamiltonians $\{H_l\mid l=0,1,\cdots,L-1\}$ which have ground states $\ket{g_l}$ that are mutually almost orthogonal, that is, $\left\langle g_{l} \mid g_{l^\prime}\right\rangle \approx 0$ for $l \neq l^{\prime}$.
Labeling each optimized ansatz $U_m$ based on the Hamiltonian to which it is optimized, we define a dataset $\{(U_m, l_m)\mid m=1,2,\cdots,M\}$ and $l_m\in\{0,1,\cdots,L-1\}$ as a set of pairs of a quantum circuit and its label.

We can expect this dataset to have a nice property that $|\bra{0}U_m^\dagger U_{m^\prime}\ket{0}|^2\approx 1$ when $l_m=l_{m^\prime}$ and $|\bra{0}U_m^\dagger U_{m^\prime}\ket{0}|^2\ll 1$ otherwise.
Note that this property persists even if the optimization is imperfect.
Suppose two quantum circuits $U_1(\bm{\theta}_1)$ and $U_2(\bm{\theta}_2)$ are respectively optimized to output non-degenerate ground states $\ket{g_{l_1}}$ and $\ket{g_{l_2}}$ of Hamiltonians $H_{l_1}$ and $H_{l_2}$.
Furthermore, consider a scenario where the optimization is imperfect, such that for instance, $\left|\left\langle g_{l_{m}}\left|U_{m}\left(\boldsymbol{\theta}_{m}\right)\right| 0\right\rangle\right|^{2}\geq\frac{3}{4}$ for $m=1,2.$
When $H_{l_1}=H_{l_2}$, $|\bra{0}U_1^\dagger(\bm{\theta}_1)U_2(\bm{\theta}_2)\ket{0}|^2\geq 1/4$. 
On the contrary, assuming $H_{l_1}\neq H_{l_2}$ and $\braket{g_{l_1}}{g_{l_2}}=0$, $|\bra{0}U_1^\dagger(\bm{\theta}_1)U_2(\bm{\theta}_2)\ket{0}|^2\leq 1/16$.
We expect this property makes it easier to extend the dataset by actual experiments using quantum hardware.

\subsection{Dataset construction details}

Table \ref{tab:overview} shows the overview of the dataset that we provide in this work.
To define the $L=6$ dataset, we use the Hamiltonians in Table \ref{tab:hamiltonians}.
$X_n$, $Y_n$ and $Z_n$ ($n=1,2,\cdots,N$) are respectively Pauli $X$, $Y$ and $Z$ operators acting on the $n$th qubit.
$a^\dagger_{i,\sigma}$ and $a_{i,\sigma}$ is fermion creation and annihilation operators acting on the $i$th site with the spin $\sigma$.
For the Hubbard models, we map the Hamiltonians to qubit ones by Jordan-Wigner transformation; specifically, defining $\tilde{a}_i$ ($i=1,2,...,N$) as $\tilde{a}_{2j-1}=a_{j,\uparrow}$ and $\tilde{a}_{2j}=a_{j,\downarrow}$ for $j=1,2,...,N/2$, we map them as
\begin{equation}
    \tilde{a}_i \to \frac{1}{2}(X_i+iY_i)Z_{i-1}\cdots Z_{1}
\end{equation}
The sites of 2D Hubbard model are defined on $1\times 2$, $2\times 2$, $3\times 2$, $4\times 2$, and $5\times 2$ square grids for $N=4$, 8, 12, 16, and 20 respectively, and a 2D site index $(j_x,j_y)$ is mapped to the one-dimensional index $j$ as $j=2(j_x-1)+j_y$ for $j_x=1,2,\cdots,N/4$ and $j_y=1,2$. 

For ansatz, we use the ones listed in Table \ref{tab:ansatz}. 
The one which we refer to as hardware-efficient ansatz has been a popular choice for the VQE ansatz to perform proof-of-principle demonstration of ideas \cite{kandala2017hardware, ssvqe, Kubler2020adaptiveoptimizer, nft, mcclean2018barren}.
The brick-block ansatz where we sequentially apply parameterized two-qubit gates are also a popular choice \cite{parrish2019, Slattery2022, Dborin_2022, Rodolph2022}.
The last type of ansatz used in this work is the Hamiltonian ansatz where we sequentially apply Pauli rotations $R_\sigma(\theta) = \exp(-i\theta \sigma/2)$ for all Pauli operators $\sigma=X,Y,Z$ appearing in the problem Hamiltonian.
This ansatz is physically motivated by the adiabatic evolution \cite{Wecker2015, Wierseme2020}.
For each ansatz, we vary the circuit depth $D$ from 3 to 32 so that we obtain 30 optimized circuits from each ansatz type.
For each $D$, we sample initial parameters for optimization from the uniform distribution on $[-2\pi, 2\pi)$, except for the Hamiltonian ansatz where we sample from $[0, 0.1)$, for ten times, and adopt the one which achieves the lowest energy expectation value after the optimization to the dataset.
The optimization of parameters is performed by the Broyden-Fletcher-Goldfarb-Shanno (BFGS) method implemented on SciPy \cite{2020SciPy-NMeth}.
Using the exact expectation values without statistical noise computed by Qulacs \cite{qulacs}, we optimize the parameters until the norm of gradient becomes less than $10^{-5}$ or the number of iterations exceeds $1000$.

\begin{table}[t]
    \caption{Overview of the dataset.\label{tab:overview}}
    \begin{ruledtabular}
        \begin{tabular}{p{0.45\linewidth}p{0.15\linewidth}p{0.4\linewidth}}
            Numbers of qubits & $N$ & $4,8,12,16,20$ \\
            Number of labels & $L$ & 6 (5 for $N=4$) \\
            Number of circuits for each label & $M/L$ & 300\\
            Total number of circuits & $M$ & 1800 (1500 for $N=4$)
        \end{tabular}
    \end{ruledtabular}
\end{table}

\begin{table*}[t]
    \caption{Hamiltonians used to generate the dataset. \label{tab:hamiltonians}}
    \begin{ruledtabular}
        \begin{tabular}{p{0.1\linewidth}p{0.25\linewidth}p{0.65\linewidth}}
        Label & Name & Hamiltonian \\
        0 & 1D transverse-field Ising model & $\sum_{n=1}^{N-1} Z_n Z_{n+1} + 2\sum_{n=1}^{N} X_n$ \\
        1 & 1D Heisenberg model & $\sum_{n=1}^{N-1} (X_n X_{n+1} + Y_n Y_{n+1} + Z_n Z_{n+1}) + 2\sum_{n=1}^{N} Z_n$ \\
        2 & Su-Schrieffer-Heeger model & $\sum_{n=1}^{N-1}\left(1+\frac{3}{2}(-1)^n\right)\left(X_n X_{n+1}+Y_n Y_{n+1}+Z_n Z_{n+1}\right)$ \\
        3 & $J_1$-$J_2$ model & $\sum_{n=1}^{N-1}\left[\left(X_n X_{n+1}+Y_n Y_{n+1}+Z_n Z_{n+1}\right)+3\left(X_n X_{n+2}+Y_n Y_{n+2}+Z_n Z_{n+2}\right)\right]$ \\
        4 & 1D Hubbard model & $-\sum_{j=1}^{N/2-1} \sum_{\sigma \in \{\uparrow,\downarrow\}}\left(a_{j, \sigma}^{\dagger} a_{j+1, \sigma}+\mathrm{H.c.}\right)+ \sum_{j=1}^{N/2}\left(a_{j, \uparrow}^\dagger a_{j, \uparrow}-\frac{1}{2}\right)\left(a_{j, \downarrow}^\dagger a_{j, \downarrow}-\frac{1}{2}\right)$ \\
        5 & 2D Hubbard model & $-\sum_{\sigma \in \{\uparrow,\downarrow\}}\left(\sum_{j_x=1}^{N/4-1}\sum_{j_y=1}^2 a_{j_x,j_y, \sigma}^{\dagger} a_{j_x+1,j_y, \sigma}+\sum_{j_x=1}^{N/4}a_{j_x,1, \sigma}^{\dagger} a_{j_x,2, \sigma}+\mathrm{H.c.}\right)$
        $+ \sum_{j_x=1}^{N/4}\sum_{j_y=1}^2\left(a_{j_x,j_y, \uparrow}^\dagger a_{j_x,j_y, \uparrow}-\frac{1}{2}\right)\left(a_{j_x,j_y, \downarrow}^\dagger a_{j_x,j_y, \downarrow}-\frac{1}{2}\right)$
        \end{tabular}
    \end{ruledtabular}
\end{table*}

\begin{table*}[t]
    \caption{Ansatz used to generate the dataset. Angles in every rotation gate are treated as independent parameters of circuit. $R_\sigma$ is a rotation gate defined as $R_\sigma(\theta) = \exp(-i\theta\sigma /2)$ for Pauli operators $\sigma\in\{I,X,Y,Z\}^{\otimes N}$. $V_{n,n^\prime}$ is a four-parameter unitary gate defined as $V_{n,n^\prime}=\mathrm{C_X}_{n,n^\prime}R_{Y_n}R_{Y_{n^\prime}}\mathrm{CNOT}_{n,n^\prime}R_{Y_n}R_{Y_{n^\prime}}$, where $\mathrm{CNOT}_{n,n^\prime}$ is a NOT gate on the qubit $n^\prime$ controlled with the qubit $n$, which is taken from \cite{parrish2019}.\label{tab:ansatz}
    In addition, the following $\mathcal{P}$s represent various ordered sets of pairs of sites for the 2-qubit gate.
    $\mathcal{P}_{\mathrm{chain}} = \{ (N-2j-1,N-2j) \mid j=1,2,\cdots,N/2-1\} \cup \{ (N-2j,N-2j+1) \mid j=1,2,\cdots,N/2\}$,
    $\mathcal{P}_{\mathrm{stair}} = \{ (N-n,N-n+1) \mid n=1,2,\cdots,N-1\}$, 
    $\mathcal{P}_{\mathrm{complete}} = \{ (N-n,N-n^\prime)\mid n^\prime=0,1,\cdots,n-1; n=1,2,\cdots, N-1 \}$, 
    $\mathcal{P}_{\mathrm{ladder}} =  \{ (N-2j-1,N-2j+1) \mid j=1,2,\cdots,N/2-1\} \cup \{ (N-2j,N-2j+2) \mid j=1,2,\cdots,N/2-1\} \cup  \{ (2j-1,2j)\mid j=1,2,\cdots,N/2 \} $, 
    $\mathcal{P}_{\mathrm{cross-ladder}} = \mathcal{P}_{\mathrm{ladder}}\cup 
    \{ (N-2j,N-2j+1),(N-2j-1,N-2j+2)\mid j=1,2,\cdots,N/2-1\}$
    , where $N$ is the number of qubits.
    } 
    \begin{ruledtabular}
        \begin{tabular}{p{0.5\linewidth}p{0.5\linewidth}}
        Name & Ansatz definition \\
        Hamiltonian ansatz & $\prod_{d=1}^D \left(\prod_{\sigma\in H} R_{\sigma}\prod_{n=1}^{N} R_{X_n}R_{Z_n} \right)$ \\
        
        Hardware-efficient ansatz (HE) & $\prod_{d=1}^D \qty[\prod_{n=1}^{N} R_{Z_n}R_{Y_n}
        \left[\prod_{p\in\mathcal{P}_{\mathrm{chain}}}\mathrm{CZ}_{p}\right]
        ] \prod_{n=1}^{N} R_{Z_n}R_{Y_n}$ \\

        Complete Hardware-efficient ansatz (Complete-HE) & $\prod_{d=1}^D \qty[\prod_{n=1}^{N} R_{Z_n}R_{Y_n} 
        \left[
        \prod_{p\in\mathcal{P}_{\mathrm{complete}}}\mathrm{CZ}_{p}
        \right]
        ]\prod_{n=1}^{N} R_{Z_n}R_{Y_n}$ \\

        Ladder Hardware-efficient ansatz (Ladder-HE) & $\prod_{d=1}^D \qty[\prod_{n=1}^{N} R_{Z_n}R_{Y_n}
        \left[\prod_{p\in\mathcal{P}_{\mathrm{ladder}}}\mathrm{CZ}_{p}\right]
        ]\prod_{n=1}^{N} R_{Z_n}R_{Y_n}$ \\

        Cross-Ladder Hardware-efficient ansatz (Cross-ladder-HE) & $\prod_{d=1}^D \qty[\prod_{n=1}^{N} R_{Z_n}R_{Y_n}
        \left[\prod_{p\in\mathcal{P}_{\mathrm{cross-ladder}}}\mathrm{CZ}_{p}\right]
        ]\prod_{n=1}^{N} R_{Z_n}R_{Y_n}$ \\
        
        1D brick-block ansatz (1D-BB) & $\prod_{n=1}^N R_{Y_n} \prod_{d=1}^D
        \left[\prod_{p\in\mathcal{P}_{\mathrm{chain}}}\mathrm{V}_{p}\right]
        $ \\

        Stair brick-block ansatz (Stair-BB) & $\prod_{n=1}^N R_{Y_n} \prod_{d=1}^D
        \left[\prod_{p\in\mathcal{P}_{\mathrm{stair}}}\mathrm{V}_{p}\right]
        $ \\

        Complete brick-block ansatz (Complete-BB) & $\prod_{n=1}^N R_{Y_n} \prod_{d=1}^D
        \left[\prod_{p\in\mathcal{P}_{\mathrm{complete}}}\mathrm{V}_{p}\right]
        $ \\
        
        Ladder brick-block ansatz (Ladder-BB) & $\prod_{n=1}^N R_{Y_n} \prod_{d=1}^D
        \left[\prod_{p\in\mathcal{P}_{\mathrm{ladder}}}\mathrm{V}_{p}\right]
        $ \\

        Cross-Ladder brick-block ansatz (Cross-ladder-BB) & $\prod_{n=1}^N R_{Y_n} \prod_{d=1}^D
         \left[\prod_{p\in\mathcal{P}_{\mathrm{cross-ladder}}}\mathrm{V}_{p}\right]
        $
        
        \end{tabular}
    \end{ruledtabular}
\end{table*}

\section{Dataset properties}

\subsection{Visualization of dataset}
First, we visualize the constructed dataset using t-stochastic neighbor embedding (t-SNE) \cite{vanDerMaaten2008} to understand the distribution of data intuitively.
t-SNE is a visualization method which takes a distance matrix $d_{m,m^\prime}$ of a dataset $\{\bm{x}_m\}$ consisting of high-dimensional vectors as its input, and generate low-dimensional points which maintain the similarities among the data points.
Here, we adopt $d_{m,m^\prime} = 1-|\bra{0}U_m^\dagger U_{m^\prime}\ket{0}|^2$ as the distance matrix of our dataset consisting of quantum circuits $\{U_m\}$.
We use the exact values for inner product of $\bra{0}U_m^\dagger U_{m^\prime}\ket{0}$ calculated by Qulacs \cite{qulacs}.

\begin{figure*}[!htbp]
  \centering
  \includegraphics[height=\linewidth]{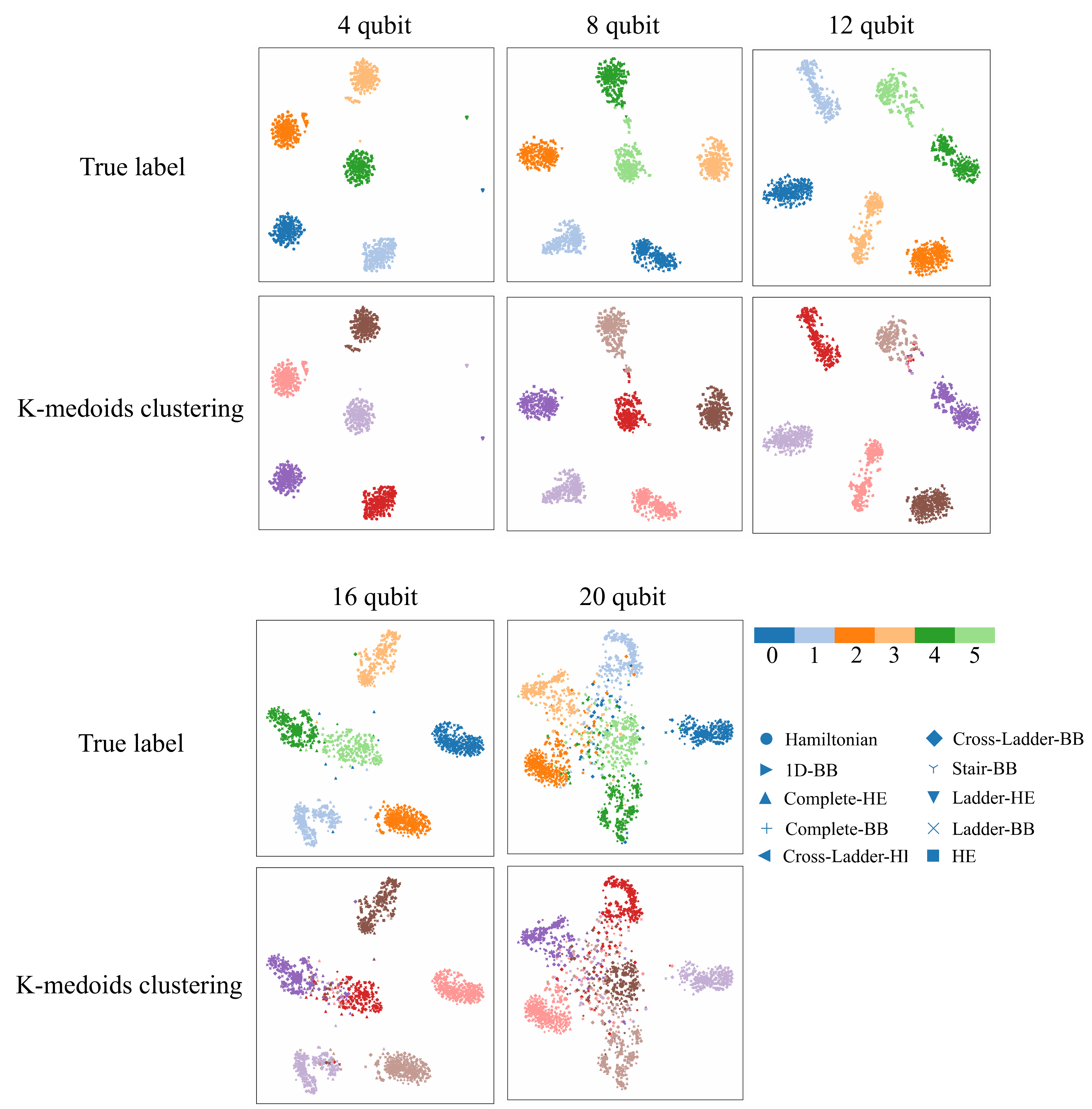}
  \caption{\textbf{Visualization and clustering result of the dataset.}
 (Top panels) Visualization of the dataset using t-SNE. The points are colored depending on their true labels. (Bottom panels) Clustering result by k-medoids algorithm. The points are colored based on the clusters. Both plots are generated using exact values of distance matrix $d_{m,m^\prime}$. 
 Different markers correspond to different ansatzes.}
  \label{fig:clustering_result}
\end{figure*}

The visualization result is shown in Fig.~\ref{fig:clustering_result} (top panels).
Each point corresponds to a circuit $U_m$, and it is colored depending on its label.
We observe that the 4-, 8-, 12-  and 16-qubit datasets are well clustered, while a portion of the data from the 20-qubit dataset appear to be somewhat intermingled.
We find it harder to solve Hamiltonians corresponding to 20 qubits than the others with some ansatz employed in this work.
These outcomes depend on the chosen distance metric.
Alternative metrics may yield disparate results.
Furthermore, the inability to achieve impeccable clustering of the dataset is not regarded as a deficiency because it allows for the comparison of different clustering algorithms.
We present in the Appendix the fidelity between the output state and the true ground state of the Hamiltonian for each ansatz.
Another interesting feature is the existence of multiple clusters in a label.
We believe this is an artificial effect caused by the t-SNE visualization.
This feature cannot be observed by another visualization technique called multidimensional scaling (MDS) \cite{Kruskal1964b}, which is shown in the Appendix.

To demonstrate the ease of clustering the proposed dataset for ideal quantum computers, we perform clustering based on the exact value of $d_{ij}$. We note that this is not the only approach for solving the proposed learning task.
We employ the $k$-medoids algorithm implemented in PyClustering\cite{Novikov2019}. 
The performance of the clustering is evaluated by adjusted Rand index (ARI) \cite{hubert1985comparing}, which takes a value between 0 (random clustering) and 1 (perfect clustering).
The ARI is evaluated as the mean of ten trials of clustering with different random seeds.

The result is visualized in Fig.~\ref{fig:clustering_result}.
The ARI of the 4-, 8-, 12-, 16- and 20-qubit dataset is respectively 0.992, 0.968, 0.927, 0.883 and 0.692.
The relatively low ARI for the 20-qubit dataset is caused by the difficulty of producing quantum circuits to output the ground state of label 5.
This result indicates that it is possible to cluster this dataset, even in the unsupervised setting, with an ideal quantum computer.

\thispagestyle{empty}
\clearpage
\subsection{Clustering using real quantum hardware and noise model simulator}
Here, we show our four-qubit dataset can be reliably learned by using real quantum computers that are presently available. 
To this end, we perform the clustering by running quantum circuits $U_m^\dagger U_{m^\prime}$ for all pairs of $m$ and $m^\prime$ on the \texttt{ibmq\_manila} device available at IBMQ to get fidelity between the two output states.
The number of measurements is set to $2\times10^4$ for each $U_m^\dagger U_{m^\prime}$.
Only hardware-efficient ansatz (HE) and 1D brick-block (1D-BB) ansatz with $D=3$ to $12$ are used in the experiments, and thus the number of data is ten for each ansatz.

\begin{figure}
 \centering
 \includegraphics[width=\linewidth]{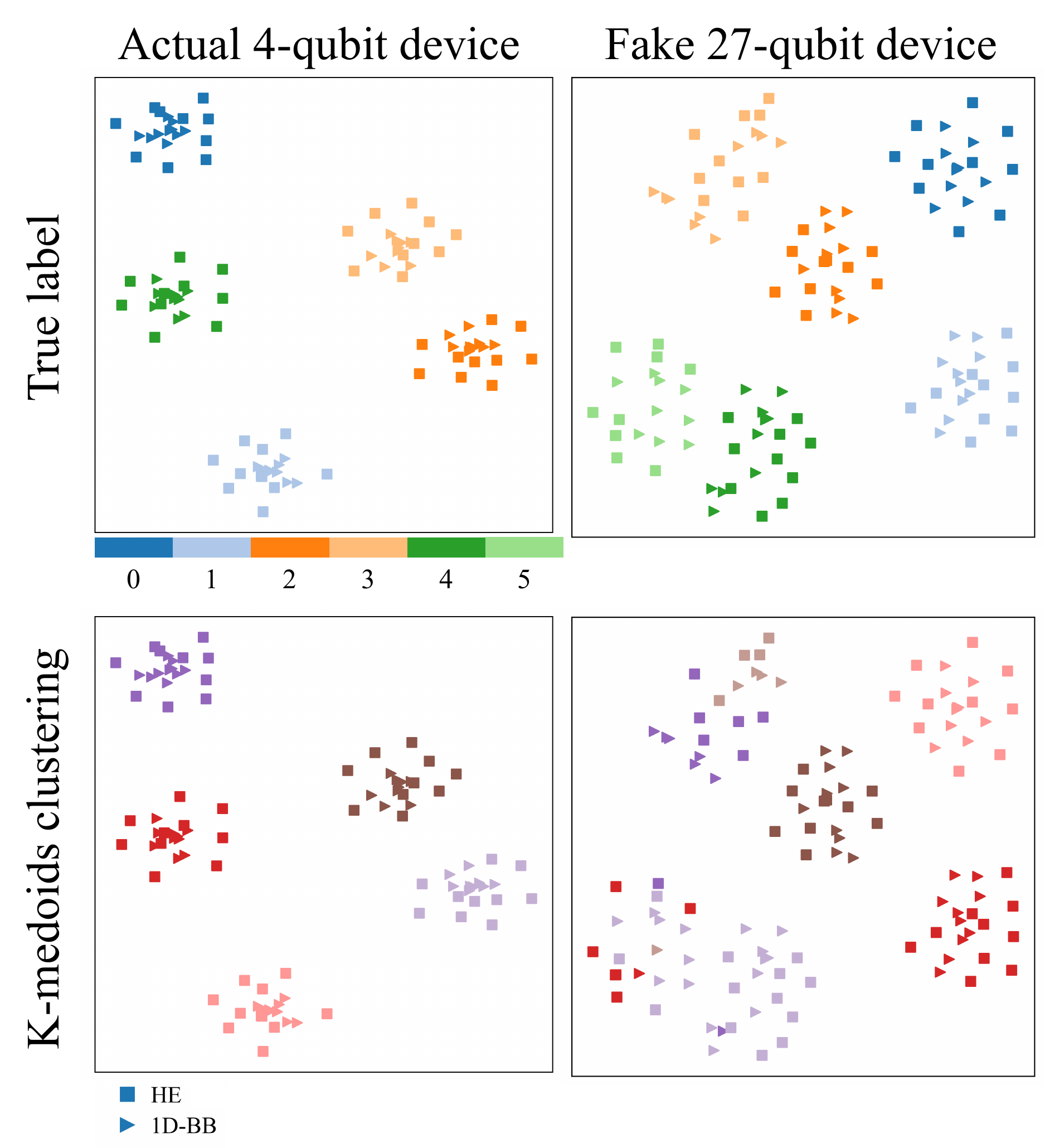}
 \caption{\textbf{Results of clustering quantum circuit in the dataset using IBM's four-qubit quantum computer and 27-qubit fake device.} 
 (Left panel) The result of clustering on the four-qubit dataset by IBM's four-qubit quantum computer.
 (right panel) The result of clustering on the 20-qubit dataset by noise model simulation using real noise data of IBM's 27-qubit quantum computer.
 (Top panel) Visualization of the dataset using t-SNE. The points are colored depending on their true labels.
 (Bottom panel) Clustering result by $k$-medoids algorithm.}
\label{fig:ibmq_result}
\end{figure}

In Fig.~\ref{fig:ibmq_result}, we visualize the dataset by t-SNE using the distance matrix obtained from the experiment.
As we can observe, we are able to perfectly cluster the dataset.
This is because fidelities between the quantum states belonging to the same labels are maintained to be much larger than those with different labels even in a noisy environment.
This result shows the actual quantum computers are capable of learning the dataset we propose.

To investigate the possibility of learning our 20-qubit dataset on actual devices, we also perform clustering by running circuits on \texttt{FakeAuckland} backend available at IBMQ which is a simulator mimicking \texttt{ibm\_auckland} device.
The result is visualized in Fig. \ref{fig:ibmq_result} in the same manner as the above experiment. 
The ARI is 0.720. 
This indicates the possibility of learning our dataset using real devices even at the size of 20 qubits.

\subsection{Classical machine learning applied to the dataset}

We expect our dataset to be hard to be learned solely by classical computers.
The most challenging aspect of applying classical machine learning algorithms is identifying an effective feature map that can transform the classical descriptions of a quantum circuit--such as gate types, gate positions, gate parameters, and the sequence of gates--into, for instance, a vector of real values, which can then be processed by various machine learning models.
We adopted two approaches:
First, we chose to represent quantum circuits as graphs, inspired by recent literature~\cite{2210.16724v1}, and applied graph neural networks (GNNs) because they are capable of leveraging the full structural information of the circuits.
Second, we directly use the circuit parameters as features and apply support vector machine (SVM) classifiers, to test how much information can be learned from a simplified representation.

\begin{figure}[h]
 \centering
 \includegraphics[width=\linewidth]{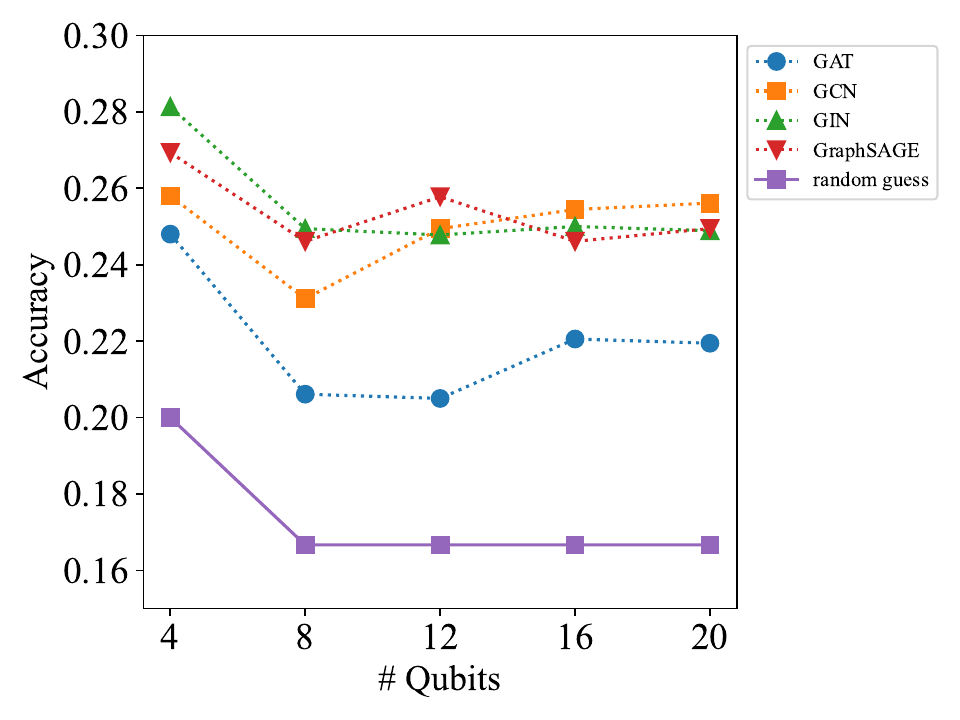}
 \caption{\textbf{Performance of different graph-based models for the classification of the dataset.}  
 }
 \label{fig:graph_result}
\end{figure}

In the former, we preprocessed the data transforming each circuit to graph.
We selected a graph structure inspired by Ref.~\cite{2210.16724v1}.
A circuit is translated to a graph whose nodes represent gates and encode information about gates as real vectors, and edges are placed basically when there is a wire in a quantum circuit diagram. 
We test four graph-based models: graph convolutional networks (GCN)~\cite{kipf2016semisupervised}, graph attention networks (GAT)~\cite{veličković2018graph}, graph isomorphism network (GIN)~\cite{xu2019powerful} and graph sample and aggregate (GraphSAGE)~\cite{hamilton2017inductive}, and estimated the accuracy with three-fold cross validation. The performance hardly reaches 30\% multiclass accuracy on the training set, showing the hardness of the task. 
The resulting performance on the validation set can be seen in Fig.~\ref{fig:graph_result}. 
For details about the models and their implementations, see ~\cite{nakayama2023github}.

In the latter, we consider the task of classifying a dataset comprised of circuits, all utilizing the same type of ansatz, by employing the circuits' parameter vector, $\bm{\theta}$, as features within a supervised learning framework. 
Here, the classical approach is restricted to use only a subset of the information available to the quantum circuits, which is the parameter vector $\bm{\theta}$.
For the model, we employ kernel SVM \cite{PlattProbabilisticOutputs1999}. 
To remove the difference in the lengths of the parameter vectors $\bm{\theta}$ depending on the different depths of the circuits, we extend $\bm{\theta}$ to match the deepest circuits within each ansatz type and fill the extended elements with zero.
We split 80\% of all data into training data and the rest into test data to ensure that the divided data has roughly the same proportions of different labels.
We train the model ten times with different splitting and report the mean accuracy score.
The regularization strength, types of kernel, and the parameters in kernel are treated as hyperparameters and optimized through the grid search technique combined with stratified threefold cross validation.
We note that other classification methods such as random forest have given us similar results.

\begin{figure}
 \centering
 \includegraphics[width=\linewidth]{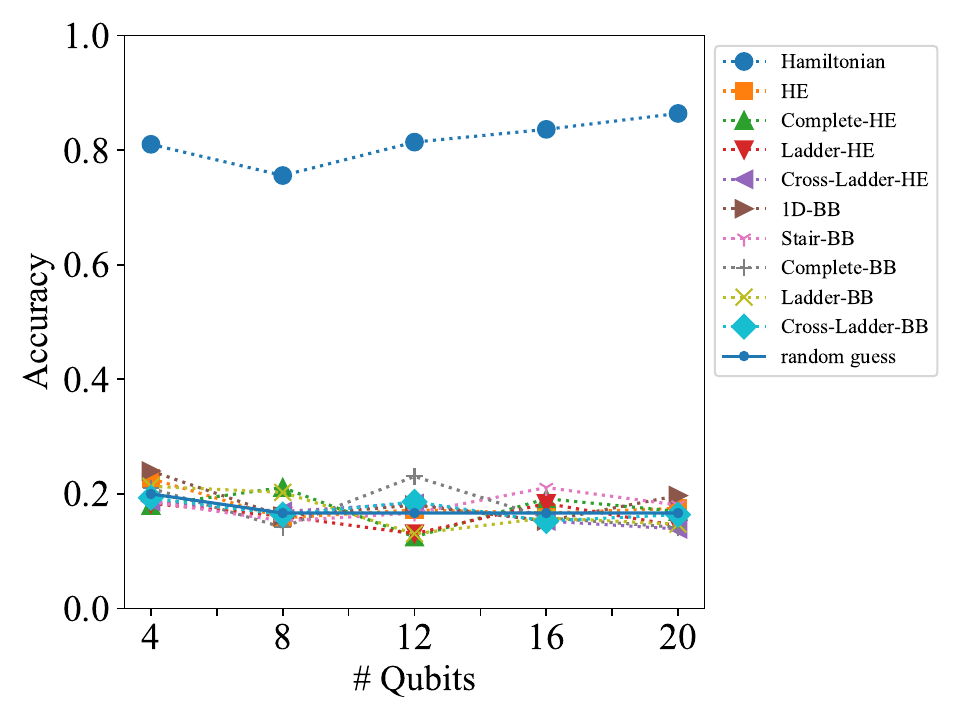}
 \caption{\textbf{Classification result using support vector machine on the circuit parameters.}  
 }
 \label{fig:svm_result}
\end{figure}

Figure~\ref{fig:svm_result} shows the classification accuracy.
For Hamiltonian ansatz, we reached a higher classification accuracy of up to about 80\% than other ans\"{a}tze for all $n$.
Such a high accuracy may be caused by the difference of quantum circuit structures from one label to another, which only exist in the case of the Hamiltonian ansatz.
Except for the Hamiltonian ansatz, we only reached a classification accuracy of up to about 20\% accuracy.
This means that labels are predicted almost randomly.
Although there may be a possible improvement to be made to this result by using more sophisticated methods such as neural networks, we can at least say “standard” classification models such as SVM do not work well on the proposed dataset even for the very simplified task.

\section{Conclusion}
In this paper, we proposed a quantum circuit classification problem as a more practical machine learning task. 
We introduced the dataset of $N=4,8,12,16,20$-qubit quantum circuits optimized by VQE for different Hamiltonians using different types of ansatz.
We verified that the unsupervised clustering of the dataset is easy for ideal quantum computers.
In 4-, 8-, 12-, and 16-qubit cases, we achieved the ARI score of over 0.88, and in the 20-qubit case, we achieved 0.69.
In particular, we demonstrate a successful classification of our four-qubit subdataset using the actual four-qubit device.

Potential future directions include the following:
First, the capability of generating datasets through VQE techniques suggests that actual quantum devices may be able to produce datasets encompassing an even greater number of qubits. This strategy is considered feasible with the careful selection of suitable ansatz and optimization methods. Pertinent research has explored various methods, including pretraining \cite{grant2019initialization}, layerwise training \cite{skolik2021layerwise}, and strategies employing tensor networks \cite{rudolph2022synergy,watanabe2023entangled}.
It is also interesting to explore whether other variational quantum algorithms or fault tolerant quantum computer algorithms can also construct a dataset similar to the one provided in this work.
Our dataset, which provides a set of practical quantum circuits optimized to output ground states, can also be useful as a benchmark for quantum circuit compilers/transpilers.
The devices may be able to conduct similar experiments on our dataset of more numbers of qubits.
We finally note that it is worth investigating whether state-of-the-art classical machine learning algorithms can solve the dataset. 
We published the dataset on GitHub \cite{nakayama2023github}.

\begin{acknowledgments}
K.M. is supported by JST PRESTO Grants No. JPMJPR2019, JST FOREST Grant No. JPMJFR232Z, JSPS KAKENHI Grant No. 20K22330, 23H03819, 24K16980 and JST CREST Grant No. JPMJCR24I4.
T.S. is supported by JST PRESTO Grant No. JPMJPR24F4.
K.F. is supported by JST ERATO Grant No. JPMJER1601 and JST CREST Grant No. JPMJCR1673.
This work is supported by MEXT Quantum Leap Flagship Program (MEXT-QLEAP) Grant No. JPMXS0120319794 and JST COI-NEXT Grant No. JPMJPF2014.
\end{acknowledgments}

\appendix*
\section{supplementary data analysis}
\begin{figure*}
 \includegraphics[width=\linewidth]{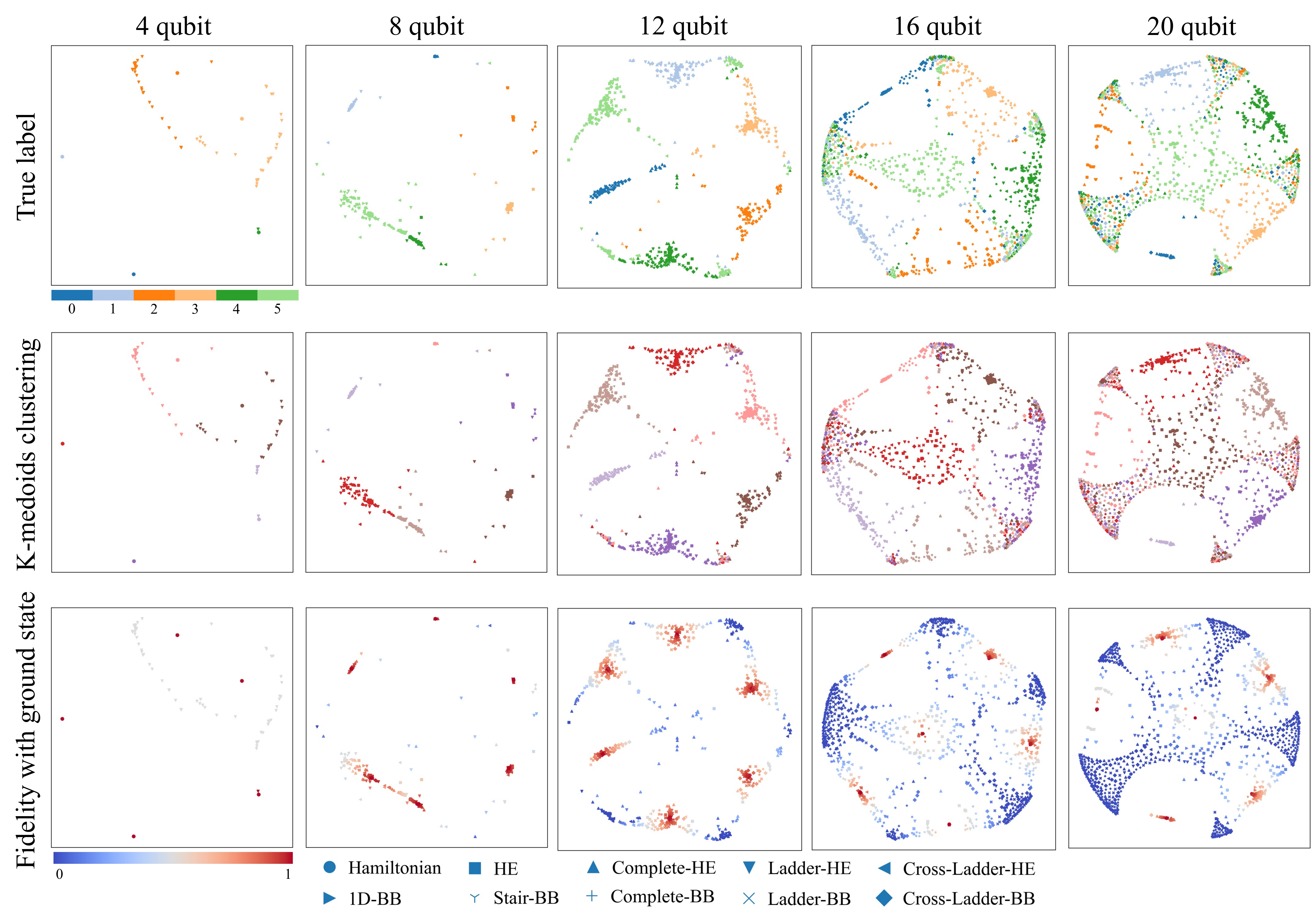}
 \caption{\textbf{Visualization and clustering result of the dataset}
  (Top panel) Visualization of the dataset using MDS. The points are colored depending on their true labels. (Middle panel) Clustering result by k-medoids algorithm.
  (Bottom panel) The same visualization but are colored depending on their fidelity with the ground state of each label's Hamiltonian.
  }
 \label{fig:result_appendix}
\end{figure*}

\begin{figure*}
 \includegraphics[width=\linewidth]{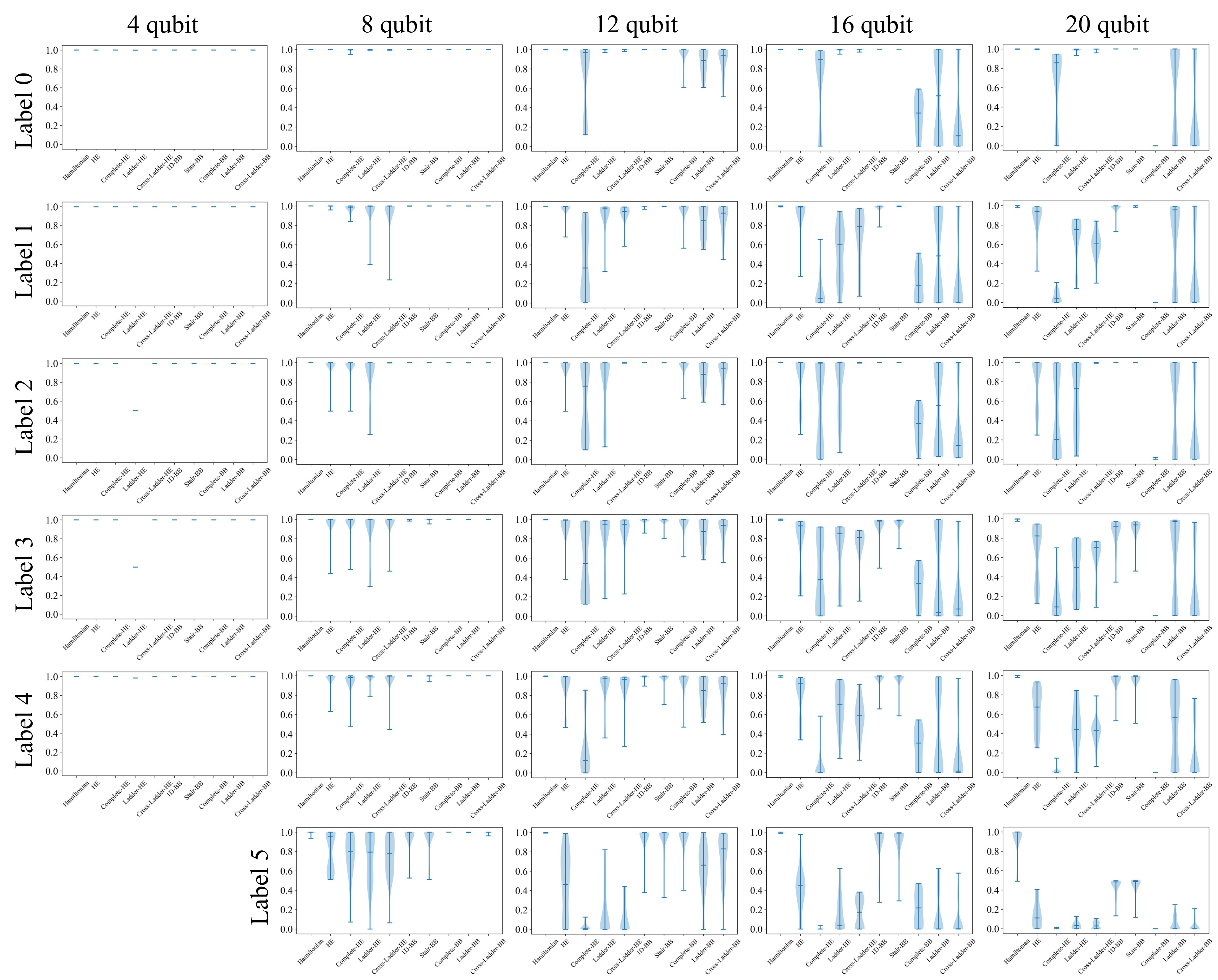}
 \caption{\textbf{Fidelity between each data and the ground state of each label's Hamiltonian}}
 \label{fig:fidelity_appendix}
\end{figure*}

Here we present additional details for the main results in our manuscript.
In Fig. \ref{fig:result_appendix}, we present the visualizations of the dataset by MDS \cite{Kruskal1964b}. 
Figure \ref{fig:fidelity_appendix} shows the fidelity between each data and ground state of the corresponding Hamiltonian by violin plots to understand how close the output of each circuit is to the ground states.

\bibliography{1_main}

\begin{thebibliography}{54}%
\makeatletter
\providecommand \@ifxundefined [1]{%
 \@ifx{#1\undefined}
}%
\providecommand \@ifnum [1]{%
 \ifnum #1\expandafter \@firstoftwo
 \else \expandafter \@secondoftwo
 \fi
}%
\providecommand \@ifx [1]{%
 \ifx #1\expandafter \@firstoftwo
 \else \expandafter \@secondoftwo
 \fi
}%
\providecommand \natexlab [1]{#1}%
\providecommand \enquote  [1]{``#1''}%
\providecommand \bibnamefont  [1]{#1}%
\providecommand \bibfnamefont [1]{#1}%
\providecommand \citenamefont [1]{#1}%
\providecommand \href@noop [0]{\@secondoftwo}%
\providecommand \href [0]{\begingroup \@sanitize@url \@href}%
\providecommand \@href[1]{\@@startlink{#1}\@@href}%
\providecommand \@@href[1]{\endgroup#1\@@endlink}%
\providecommand \@sanitize@url [0]{\catcode `\\12\catcode `\$12\catcode
  `\&12\catcode `\#12\catcode `\^12\catcode `\_12\catcode `\%12\relax}%
\providecommand \@@startlink[1]{}%
\providecommand \@@endlink[0]{}%
\providecommand \url  [0]{\begingroup\@sanitize@url \@url }%
\providecommand \@url [1]{\endgroup\@href {#1}{\urlprefix }}%
\providecommand \urlprefix  [0]{URL }%
\providecommand \Eprint [0]{\href }%
\providecommand \doibase [0]{https://doi.org/}%
\providecommand \selectlanguage [0]{\@gobble}%
\providecommand \bibinfo  [0]{\@secondoftwo}%
\providecommand \bibfield  [0]{\@secondoftwo}%
\providecommand \translation [1]{[#1]}%
\providecommand \BibitemOpen [0]{}%
\providecommand \bibitemStop [0]{}%
\providecommand \bibitemNoStop [0]{.\EOS\space}%
\providecommand \EOS [0]{\spacefactor3000\relax}%
\providecommand \BibitemShut  [1]{\csname bibitem#1\endcsname}%
\let\auto@bib@innerbib\@empty
\bibitem [{\citenamefont {Biamonte}\ \emph {et~al.}(2017)\citenamefont
  {Biamonte}, \citenamefont {Wittek}, \citenamefont {Pancotti}, \citenamefont
  {Rebentrost}, \citenamefont {Wiebe},\ and\ \citenamefont
  {Lloyd}}]{biamonte2017quantum}%
  \BibitemOpen
  \bibfield  {author} {\bibinfo {author} {\bibfnamefont {J.}~\bibnamefont
  {Biamonte}}, \bibinfo {author} {\bibfnamefont {P.}~\bibnamefont {Wittek}},
  \bibinfo {author} {\bibfnamefont {N.}~\bibnamefont {Pancotti}}, \bibinfo
  {author} {\bibfnamefont {P.}~\bibnamefont {Rebentrost}}, \bibinfo {author}
  {\bibfnamefont {N.}~\bibnamefont {Wiebe}},\ and\ \bibinfo {author}
  {\bibfnamefont {S.}~\bibnamefont {Lloyd}},\ }\bibfield  {title} {\bibinfo
  {title} {Quantum machine learning},\ }\href
  {https://doi.org/10.1038/nature23474} {\bibfield  {journal} {\bibinfo
  {journal} {Nature}\ }\textbf {\bibinfo {volume} {549}},\ \bibinfo {pages}
  {195} (\bibinfo {year} {2017})}\BibitemShut {NoStop}%
\bibitem [{\citenamefont {Cerezo}\ \emph {et~al.}(2022)\citenamefont {Cerezo},
  \citenamefont {Verdon}, \citenamefont {Huang}, \citenamefont {Cincio},\ and\
  \citenamefont {Coles}}]{cerezo2022challenges}%
  \BibitemOpen
  \bibfield  {author} {\bibinfo {author} {\bibfnamefont {M.}~\bibnamefont
  {Cerezo}}, \bibinfo {author} {\bibfnamefont {G.}~\bibnamefont {Verdon}},
  \bibinfo {author} {\bibfnamefont {H.-Y.}\ \bibnamefont {Huang}}, \bibinfo
  {author} {\bibfnamefont {L.}~\bibnamefont {Cincio}},\ and\ \bibinfo {author}
  {\bibfnamefont {P.~J.}\ \bibnamefont {Coles}},\ }\bibfield  {title} {\bibinfo
  {title} {Challenges and opportunities in quantum machine learning},\ }\href
  {https://doi.org/10.1038/s43588-022-00311-3} {\bibfield  {journal} {\bibinfo
  {journal} {Nature Computational Science}\ }\textbf {\bibinfo {volume} {2}},\
  \bibinfo {pages} {567} (\bibinfo {year} {2022})}\BibitemShut {NoStop}%
\bibitem [{\citenamefont {Farhi}\ and\ \citenamefont
  {Neven}(2018)}]{farhi2018}%
  \BibitemOpen
  \bibfield  {author} {\bibinfo {author} {\bibfnamefont {E.}~\bibnamefont
  {Farhi}}\ and\ \bibinfo {author} {\bibfnamefont {H.}~\bibnamefont {Neven}},\
  }\bibfield  {title} {\bibinfo {title} {Classification with quantum neural
  networks on near term processors},\ }\Eprint
  {https://arxiv.org/abs/1802.06002v2} {arXiv:1802.06002v2 [quant-ph]}
  (\bibinfo {year} {2018})\BibitemShut {NoStop}%
\bibitem [{\citenamefont {Mitarai}\ \emph {et~al.}(2018)\citenamefont
  {Mitarai}, \citenamefont {Negoro}, \citenamefont {Kitagawa},\ and\
  \citenamefont {Fujii}}]{mitarai2018quantum}%
  \BibitemOpen
  \bibfield  {author} {\bibinfo {author} {\bibfnamefont {K.}~\bibnamefont
  {Mitarai}}, \bibinfo {author} {\bibfnamefont {M.}~\bibnamefont {Negoro}},
  \bibinfo {author} {\bibfnamefont {M.}~\bibnamefont {Kitagawa}},\ and\
  \bibinfo {author} {\bibfnamefont {K.}~\bibnamefont {Fujii}},\ }\bibfield
  {title} {\bibinfo {title} {Quantum circuit learning},\ }\href
  {https://link.aps.org/doi/10.1103/PhysRevA.98.032309} {\bibfield  {journal}
  {\bibinfo  {journal} {Physical Review A}\ }\textbf {\bibinfo {volume} {98}},\
  \bibinfo {pages} {032309} (\bibinfo {year} {2018})}\BibitemShut {NoStop}%
\bibitem [{\citenamefont {Schuld}\ and\ \citenamefont
  {Killoran}(2019)}]{Schuld2019}%
  \BibitemOpen
  \bibfield  {author} {\bibinfo {author} {\bibfnamefont {M.}~\bibnamefont
  {Schuld}}\ and\ \bibinfo {author} {\bibfnamefont {N.}~\bibnamefont
  {Killoran}},\ }\bibfield  {title} {\bibinfo {title} {Quantum machine learning
  in feature hilbert spaces},\ }\href
  {https://doi.org/10.1103/PhysRevLett.122.040504} {\bibfield  {journal}
  {\bibinfo  {journal} {Phys. Rev. Lett.}\ }\textbf {\bibinfo {volume} {122}},\
  \bibinfo {pages} {040504} (\bibinfo {year} {2019})}\BibitemShut {NoStop}%
\bibitem [{\citenamefont {Liu}\ and\ \citenamefont {Wang}(2018)}]{Liu2018}%
  \BibitemOpen
  \bibfield  {author} {\bibinfo {author} {\bibfnamefont {J.-G.}\ \bibnamefont
  {Liu}}\ and\ \bibinfo {author} {\bibfnamefont {L.}~\bibnamefont {Wang}},\
  }\bibfield  {title} {\bibinfo {title} {Differentiable learning of quantum
  circuit born machines},\ }\href {https://doi.org/10.1103/PhysRevA.98.062324}
  {\bibfield  {journal} {\bibinfo  {journal} {Phys. Rev. A}\ }\textbf {\bibinfo
  {volume} {98}},\ \bibinfo {pages} {062324} (\bibinfo {year}
  {2018})}\BibitemShut {NoStop}%
\bibitem [{\citenamefont {Benedetti}\ \emph {et~al.}(2019)\citenamefont
  {Benedetti}, \citenamefont {Garcia-Pintos}, \citenamefont {Perdomo},
  \citenamefont {Leyton-Ortega}, \citenamefont {Nam},\ and\ \citenamefont
  {Perdomo-Ortiz}}]{Benedetti2019}%
  \BibitemOpen
  \bibfield  {author} {\bibinfo {author} {\bibfnamefont {M.}~\bibnamefont
  {Benedetti}}, \bibinfo {author} {\bibfnamefont {D.}~\bibnamefont
  {Garcia-Pintos}}, \bibinfo {author} {\bibfnamefont {O.}~\bibnamefont
  {Perdomo}}, \bibinfo {author} {\bibfnamefont {V.}~\bibnamefont
  {Leyton-Ortega}}, \bibinfo {author} {\bibfnamefont {Y.}~\bibnamefont {Nam}},\
  and\ \bibinfo {author} {\bibfnamefont {A.}~\bibnamefont {Perdomo-Ortiz}},\
  }\bibfield  {title} {\bibinfo {title} {A generative modeling approach for
  benchmarking and training shallow quantum circuits},\ }\href
  {https://doi.org/10.1038/s41534-019-0157-8} {\bibfield  {journal} {\bibinfo
  {journal} {npj Quantum Information}\ }\textbf {\bibinfo {volume} {5}},\
  \bibinfo {pages} {45} (\bibinfo {year} {2019})}\BibitemShut {NoStop}%
\bibitem [{\citenamefont {Havl{\'\i}{\v{c}}ek}\ \emph
  {et~al.}(2019)\citenamefont {Havl{\'\i}{\v{c}}ek}, \citenamefont
  {C{\'o}rcoles}, \citenamefont {Temme}, \citenamefont {Harrow}, \citenamefont
  {Kandala}, \citenamefont {Chow},\ and\ \citenamefont
  {Gambetta}}]{havlivcek2019supervised}%
  \BibitemOpen
  \bibfield  {author} {\bibinfo {author} {\bibfnamefont {V.}~\bibnamefont
  {Havl{\'\i}{\v{c}}ek}}, \bibinfo {author} {\bibfnamefont {A.~D.}\
  \bibnamefont {C{\'o}rcoles}}, \bibinfo {author} {\bibfnamefont
  {K.}~\bibnamefont {Temme}}, \bibinfo {author} {\bibfnamefont {A.~W.}\
  \bibnamefont {Harrow}}, \bibinfo {author} {\bibfnamefont {A.}~\bibnamefont
  {Kandala}}, \bibinfo {author} {\bibfnamefont {J.~M.}\ \bibnamefont {Chow}},\
  and\ \bibinfo {author} {\bibfnamefont {J.~M.}\ \bibnamefont {Gambetta}},\
  }\bibfield  {title} {\bibinfo {title} {Supervised learning with
  quantum-enhanced feature spaces},\ }\href@noop {} {\bibfield  {journal}
  {\bibinfo  {journal} {Nature}\ }\textbf {\bibinfo {volume} {567}},\ \bibinfo
  {pages} {209} (\bibinfo {year} {2019})}\BibitemShut {NoStop}%
\bibitem [{\citenamefont {Kusumoto}\ \emph {et~al.}(2021)\citenamefont
  {Kusumoto}, \citenamefont {Mitarai}, \citenamefont {Fujii}, \citenamefont
  {Kitagawa},\ and\ \citenamefont {Negoro}}]{Kusumoto2021}%
  \BibitemOpen
  \bibfield  {author} {\bibinfo {author} {\bibfnamefont {T.}~\bibnamefont
  {Kusumoto}}, \bibinfo {author} {\bibfnamefont {K.}~\bibnamefont {Mitarai}},
  \bibinfo {author} {\bibfnamefont {K.}~\bibnamefont {Fujii}}, \bibinfo
  {author} {\bibfnamefont {M.}~\bibnamefont {Kitagawa}},\ and\ \bibinfo
  {author} {\bibfnamefont {M.}~\bibnamefont {Negoro}},\ }\bibfield  {title}
  {\bibinfo {title} {Experimental quantum kernel trick with nuclear spins in a
  solid},\ }\href {https://doi.org/10.1038/s41534-021-00423-0} {\bibfield
  {journal} {\bibinfo  {journal} {npj Quantum Information}\ }\textbf {\bibinfo
  {volume} {7}},\ \bibinfo {pages} {94} (\bibinfo {year} {2021})}\BibitemShut
  {NoStop}%
\bibitem [{\citenamefont {Bartkiewicz}\ \emph {et~al.}(2020)\citenamefont
  {Bartkiewicz}, \citenamefont {Gneiting}, \citenamefont {{\v{C}}ernoch},
  \citenamefont {Jir{\'a}kov{\'a}}, \citenamefont {Lemr},\ and\ \citenamefont
  {Nori}}]{bartkiewicz2020experimental}%
  \BibitemOpen
  \bibfield  {author} {\bibinfo {author} {\bibfnamefont {K.}~\bibnamefont
  {Bartkiewicz}}, \bibinfo {author} {\bibfnamefont {C.}~\bibnamefont
  {Gneiting}}, \bibinfo {author} {\bibfnamefont {A.}~\bibnamefont
  {{\v{C}}ernoch}}, \bibinfo {author} {\bibfnamefont {K.}~\bibnamefont
  {Jir{\'a}kov{\'a}}}, \bibinfo {author} {\bibfnamefont {K.}~\bibnamefont
  {Lemr}},\ and\ \bibinfo {author} {\bibfnamefont {F.}~\bibnamefont {Nori}},\
  }\bibfield  {title} {\bibinfo {title} {Experimental kernel-based quantum
  machine learning in finite feature space},\ }\href
  {https://doi.org/10.1038/s41598-020-68911-5} {\bibfield  {journal} {\bibinfo
  {journal} {Scientific Reports}\ }\textbf {\bibinfo {volume} {10}},\ \bibinfo
  {pages} {1} (\bibinfo {year} {2020})}\BibitemShut {NoStop}%
\bibitem [{\citenamefont {Rudolph}\ \emph
  {et~al.}(2022{\natexlab{a}})\citenamefont {Rudolph}, \citenamefont
  {Toussaint}, \citenamefont {Katabarwa}, \citenamefont {Johri}, \citenamefont
  {Peropadre},\ and\ \citenamefont {Perdomo-Ortiz}}]{PhysRevX.12.031010}%
  \BibitemOpen
  \bibfield  {author} {\bibinfo {author} {\bibfnamefont {M.~S.}\ \bibnamefont
  {Rudolph}}, \bibinfo {author} {\bibfnamefont {N.~B.}\ \bibnamefont
  {Toussaint}}, \bibinfo {author} {\bibfnamefont {A.}~\bibnamefont
  {Katabarwa}}, \bibinfo {author} {\bibfnamefont {S.}~\bibnamefont {Johri}},
  \bibinfo {author} {\bibfnamefont {B.}~\bibnamefont {Peropadre}},\ and\
  \bibinfo {author} {\bibfnamefont {A.}~\bibnamefont {Perdomo-Ortiz}},\
  }\bibfield  {title} {\bibinfo {title} {Generation of high-resolution
  handwritten digits with an ion-trap quantum computer},\ }\href
  {https://doi.org/10.1103/PhysRevX.12.031010} {\bibfield  {journal} {\bibinfo
  {journal} {Phys. Rev. X}\ }\textbf {\bibinfo {volume} {12}},\ \bibinfo
  {pages} {031010} (\bibinfo {year} {2022}{\natexlab{a}})}\BibitemShut
  {NoStop}%
\bibitem [{\citenamefont {Liu}\ \emph {et~al.}(2021)\citenamefont {Liu},
  \citenamefont {Arunachalam},\ and\ \citenamefont {Temme}}]{Liu_2021}%
  \BibitemOpen
  \bibfield  {author} {\bibinfo {author} {\bibfnamefont {Y.}~\bibnamefont
  {Liu}}, \bibinfo {author} {\bibfnamefont {S.}~\bibnamefont {Arunachalam}},\
  and\ \bibinfo {author} {\bibfnamefont {K.}~\bibnamefont {Temme}},\ }\bibfield
   {title} {\bibinfo {title} {A rigorous and robust quantum speed-up in
  supervised machine learning},\ }\href
  {https://doi.org/10.1038/s41567-021-01287-z} {\bibfield  {journal} {\bibinfo
  {journal} {Nature Physics}\ }\textbf {\bibinfo {volume} {17}},\ \bibinfo
  {pages} {1013} (\bibinfo {year} {2021})}\BibitemShut {NoStop}%
\bibitem [{\citenamefont {Huang}\ \emph {et~al.}(2021)\citenamefont {Huang},
  \citenamefont {Broughton}, \citenamefont {Mohseni}, \citenamefont {Babbush},
  \citenamefont {Boixo}, \citenamefont {Neven},\ and\ \citenamefont
  {McClean}}]{Huang2021}%
  \BibitemOpen
  \bibfield  {author} {\bibinfo {author} {\bibfnamefont {H.-Y.}\ \bibnamefont
  {Huang}}, \bibinfo {author} {\bibfnamefont {M.}~\bibnamefont {Broughton}},
  \bibinfo {author} {\bibfnamefont {M.}~\bibnamefont {Mohseni}}, \bibinfo
  {author} {\bibfnamefont {R.}~\bibnamefont {Babbush}}, \bibinfo {author}
  {\bibfnamefont {S.}~\bibnamefont {Boixo}}, \bibinfo {author} {\bibfnamefont
  {H.}~\bibnamefont {Neven}},\ and\ \bibinfo {author} {\bibfnamefont {J.~R.}\
  \bibnamefont {McClean}},\ }\bibfield  {title} {\bibinfo {title} {Power of
  data in quantum machine learning},\ }\href
  {https://doi.org/10.1038/s41467-021-22539-9} {\bibfield  {journal} {\bibinfo
  {journal} {Nature Communications}\ }\textbf {\bibinfo {volume} {12}},\
  \bibinfo {pages} {2631} (\bibinfo {year} {2021})}\BibitemShut {NoStop}%
\bibitem [{\citenamefont {Dunjko}\ \emph {et~al.}(2017)\citenamefont {Dunjko},
  \citenamefont {Liu}, \citenamefont {Wu},\ and\ \citenamefont
  {Taylor}}]{dunjko2017}%
  \BibitemOpen
  \bibfield  {author} {\bibinfo {author} {\bibfnamefont {V.}~\bibnamefont
  {Dunjko}}, \bibinfo {author} {\bibfnamefont {Y.-K.}\ \bibnamefont {Liu}},
  \bibinfo {author} {\bibfnamefont {X.}~\bibnamefont {Wu}},\ and\ \bibinfo
  {author} {\bibfnamefont {J.~M.}\ \bibnamefont {Taylor}},\ }\bibfield  {title}
  {\bibinfo {title} {Exponential improvements for quantum-accessible
  reinforcement learning},\ }\Eprint {https://arxiv.org/abs/1710.11160v3}
  {arXiv:1710.11160v3 [quant-ph]}  (\bibinfo {year} {2017})\BibitemShut
  {NoStop}%
\bibitem [{\citenamefont {Jerbi}\ \emph {et~al.}(2021)\citenamefont {Jerbi},
  \citenamefont {Gyurik}, \citenamefont {Marshall}, \citenamefont {Briegel},\
  and\ \citenamefont {Dunjko}}]{NEURIPS2021_eec96a7f}%
  \BibitemOpen
  \bibfield  {author} {\bibinfo {author} {\bibfnamefont {S.}~\bibnamefont
  {Jerbi}}, \bibinfo {author} {\bibfnamefont {C.}~\bibnamefont {Gyurik}},
  \bibinfo {author} {\bibfnamefont {S.}~\bibnamefont {Marshall}}, \bibinfo
  {author} {\bibfnamefont {H.}~\bibnamefont {Briegel}},\ and\ \bibinfo {author}
  {\bibfnamefont {V.}~\bibnamefont {Dunjko}},\ }\bibfield  {title} {\bibinfo
  {title} {Parametrized quantum policies for reinforcement learning},\ }in\
  \href
  {https://proceedings.neurips.cc/paper/2021/file/eec96a7f788e88184c0e713456026f3f-Paper.pdf}
  {\emph {\bibinfo {booktitle} {Advances in Neural Information Processing
  Systems}}},\ Vol.~\bibinfo {volume} {34},\ \bibinfo {editor} {edited by\
  \bibinfo {editor} {\bibfnamefont {M.}~\bibnamefont {Ranzato}}, \bibinfo
  {editor} {\bibfnamefont {A.}~\bibnamefont {Beygelzimer}}, \bibinfo {editor}
  {\bibfnamefont {Y.}~\bibnamefont {Dauphin}}, \bibinfo {editor} {\bibfnamefont
  {P.}~\bibnamefont {Liang}},\ and\ \bibinfo {editor} {\bibfnamefont {J.~W.}\
  \bibnamefont {Vaughan}}}\ (\bibinfo  {publisher} {Curran Associates, Inc.},\
  \bibinfo {year} {2021})\ pp.\ \bibinfo {pages} {28362--28375}\BibitemShut
  {NoStop}%
\bibitem [{\citenamefont {Pirnay}\ \emph {et~al.}(2023)\citenamefont {Pirnay},
  \citenamefont {Sweke}, \citenamefont {Eisert},\ and\ \citenamefont
  {Seifert}}]{Pirnay2022}%
  \BibitemOpen
  \bibfield  {author} {\bibinfo {author} {\bibfnamefont {N.}~\bibnamefont
  {Pirnay}}, \bibinfo {author} {\bibfnamefont {R.}~\bibnamefont {Sweke}},
  \bibinfo {author} {\bibfnamefont {J.}~\bibnamefont {Eisert}},\ and\ \bibinfo
  {author} {\bibfnamefont {J.-P.}\ \bibnamefont {Seifert}},\ }\bibfield
  {title} {\bibinfo {title} {Superpolynomial quantum-classical separation for
  density modeling},\ }\href@noop {} {\bibfield  {journal} {\bibinfo  {journal}
  {Physical Review A}\ }\textbf {\bibinfo {volume} {107}},\ \bibinfo {pages}
  {042416} (\bibinfo {year} {2023})}\BibitemShut {NoStop}%
\bibitem [{\citenamefont {Lecun}\ \emph {et~al.}(1998)\citenamefont {Lecun},
  \citenamefont {Bottou}, \citenamefont {Bengio},\ and\ \citenamefont
  {Haffner}}]{Lecun1998}%
  \BibitemOpen
  \bibfield  {author} {\bibinfo {author} {\bibfnamefont {Y.}~\bibnamefont
  {Lecun}}, \bibinfo {author} {\bibfnamefont {L.}~\bibnamefont {Bottou}},
  \bibinfo {author} {\bibfnamefont {Y.}~\bibnamefont {Bengio}},\ and\ \bibinfo
  {author} {\bibfnamefont {P.}~\bibnamefont {Haffner}},\ }\bibfield  {title}
  {\bibinfo {title} {Gradient-based learning applied to document recognition},\
  }\href {https://doi.org/10.1109/5.726791} {\bibfield  {journal} {\bibinfo
  {journal} {Proceedings of the IEEE}\ }\textbf {\bibinfo {volume} {86}},\
  \bibinfo {pages} {2278} (\bibinfo {year} {1998})}\BibitemShut {NoStop}%
\bibitem [{\citenamefont {Xiao}\ \emph {et~al.}(2017)\citenamefont {Xiao},
  \citenamefont {Rasul},\ and\ \citenamefont {Vollgraf}}]{fashionmnist}%
  \BibitemOpen
  \bibfield  {author} {\bibinfo {author} {\bibfnamefont {H.}~\bibnamefont
  {Xiao}}, \bibinfo {author} {\bibfnamefont {K.}~\bibnamefont {Rasul}},\ and\
  \bibinfo {author} {\bibfnamefont {R.}~\bibnamefont {Vollgraf}},\ }\bibfield
  {title} {\bibinfo {title} {Fashion-mnist: a novel image dataset for
  benchmarking machine learning algorithms},\ }\Eprint
  {https://arxiv.org/abs/1708.07747v2} {arXiv:1708.07747v2 [cs.LG]}  (\bibinfo
  {year} {2017})\BibitemShut {NoStop}%
\bibitem [{\citenamefont {Huang}\ \emph {et~al.}(2022)\citenamefont {Huang},
  \citenamefont {Broughton}, \citenamefont {Cotler}, \citenamefont {Chen},
  \citenamefont {Li}, \citenamefont {Mohseni}, \citenamefont {Neven},
  \citenamefont {Babbush}, \citenamefont {Kueng}, \citenamefont {Preskill},\
  and\ \citenamefont {McClean}}]{Huang2022}%
  \BibitemOpen
  \bibfield  {author} {\bibinfo {author} {\bibfnamefont {H.-Y.}\ \bibnamefont
  {Huang}}, \bibinfo {author} {\bibfnamefont {M.}~\bibnamefont {Broughton}},
  \bibinfo {author} {\bibfnamefont {J.}~\bibnamefont {Cotler}}, \bibinfo
  {author} {\bibfnamefont {S.}~\bibnamefont {Chen}}, \bibinfo {author}
  {\bibfnamefont {J.}~\bibnamefont {Li}}, \bibinfo {author} {\bibfnamefont
  {M.}~\bibnamefont {Mohseni}}, \bibinfo {author} {\bibfnamefont
  {H.}~\bibnamefont {Neven}}, \bibinfo {author} {\bibfnamefont
  {R.}~\bibnamefont {Babbush}}, \bibinfo {author} {\bibfnamefont
  {R.}~\bibnamefont {Kueng}}, \bibinfo {author} {\bibfnamefont
  {J.}~\bibnamefont {Preskill}},\ and\ \bibinfo {author} {\bibfnamefont
  {J.~R.}\ \bibnamefont {McClean}},\ }\bibfield  {title} {\bibinfo {title}
  {Quantum advantage in learning from experiments},\ }\href
  {https://doi.org/10.1126/science.abn7293} {\bibfield  {journal} {\bibinfo
  {journal} {Science}\ }\textbf {\bibinfo {volume} {376}},\ \bibinfo {pages}
  {1182} (\bibinfo {year} {2022})}\BibitemShut {NoStop}%
\bibitem [{\citenamefont {Schatzki}\ \emph {et~al.}(2021)\citenamefont
  {Schatzki}, \citenamefont {Arrasmith}, \citenamefont {Coles},\ and\
  \citenamefont {Cerezo}}]{ntangled}%
  \BibitemOpen
  \bibfield  {author} {\bibinfo {author} {\bibfnamefont {L.}~\bibnamefont
  {Schatzki}}, \bibinfo {author} {\bibfnamefont {A.}~\bibnamefont {Arrasmith}},
  \bibinfo {author} {\bibfnamefont {P.~J.}\ \bibnamefont {Coles}},\ and\
  \bibinfo {author} {\bibfnamefont {M.}~\bibnamefont {Cerezo}},\ }\bibfield
  {title} {\bibinfo {title} {Entangled datasets for quantum machine learning},\
  }\Eprint {https://arxiv.org/abs/2109.03400v2} {arXiv:2109.03400v2 [quant-ph]}
   (\bibinfo {year} {2021})\BibitemShut {NoStop}%
\bibitem [{\citenamefont {Perrier}\ \emph {et~al.}(2022)\citenamefont
  {Perrier}, \citenamefont {Youssry},\ and\ \citenamefont
  {Ferrie}}]{Perrier2022}%
  \BibitemOpen
  \bibfield  {author} {\bibinfo {author} {\bibfnamefont {E.}~\bibnamefont
  {Perrier}}, \bibinfo {author} {\bibfnamefont {A.}~\bibnamefont {Youssry}},\
  and\ \bibinfo {author} {\bibfnamefont {C.}~\bibnamefont {Ferrie}},\
  }\bibfield  {title} {\bibinfo {title} {Qdataset, quantum datasets for machine
  learning},\ }\href {https://doi.org/10.1038/s41597-022-01639-1} {\bibfield
  {journal} {\bibinfo  {journal} {Scientific Data}\ }\textbf {\bibinfo {volume}
  {9}},\ \bibinfo {pages} {582} (\bibinfo {year} {2022})}\BibitemShut {NoStop}%
\bibitem [{\citenamefont {Placidi}\ \emph {et~al.}(2023)\citenamefont
  {Placidi}, \citenamefont {Hataya}, \citenamefont {Mori}, \citenamefont
  {Aoyama}, \citenamefont {Morisaki}, \citenamefont {Mitarai},\ and\
  \citenamefont {Fujii}}]{placidi2023mnisq}%
  \BibitemOpen
  \bibfield  {author} {\bibinfo {author} {\bibfnamefont {L.}~\bibnamefont
  {Placidi}}, \bibinfo {author} {\bibfnamefont {R.}~\bibnamefont {Hataya}},
  \bibinfo {author} {\bibfnamefont {T.}~\bibnamefont {Mori}}, \bibinfo {author}
  {\bibfnamefont {K.}~\bibnamefont {Aoyama}}, \bibinfo {author} {\bibfnamefont
  {H.}~\bibnamefont {Morisaki}}, \bibinfo {author} {\bibfnamefont
  {K.}~\bibnamefont {Mitarai}},\ and\ \bibinfo {author} {\bibfnamefont
  {K.}~\bibnamefont {Fujii}},\ }\bibfield  {title} {\bibinfo {title} {Mnisq: A
  large-scale quantum circuit dataset for machine learning on/for quantum
  computers in the nisq era},\ }\href@noop {} {\bibfield  {journal} {\bibinfo
  {journal} {arXiv preprint arXiv:2306.16627}\ } (\bibinfo {year}
  {2023})}\BibitemShut {NoStop}%
\bibitem [{\citenamefont {Bernstein}\ and\ \citenamefont
  {Vazirani}(1997)}]{doi:10.1137/S0097539796300921}%
  \BibitemOpen
  \bibfield  {author} {\bibinfo {author} {\bibfnamefont {E.}~\bibnamefont
  {Bernstein}}\ and\ \bibinfo {author} {\bibfnamefont {U.}~\bibnamefont
  {Vazirani}},\ }\bibfield  {title} {\bibinfo {title} {Quantum complexity
  theory},\ }\href {https://doi.org/10.1137/S0097539796300921} {\bibfield
  {journal} {\bibinfo  {journal} {SIAM Journal on Computing}\ }\textbf
  {\bibinfo {volume} {26}},\ \bibinfo {pages} {1411} (\bibinfo {year}
  {1997})},\ \Eprint
  {https://arxiv.org/abs/https://doi.org/10.1137/S0097539796300921}
  {https://doi.org/10.1137/S0097539796300921} \BibitemShut {NoStop}%
\bibitem [{\citenamefont {Peruzzo}\ \emph {et~al.}(2014)\citenamefont
  {Peruzzo}, \citenamefont {McClean}, \citenamefont {Shadbolt}, \citenamefont
  {Yung}, \citenamefont {Zhou}, \citenamefont {Love}, \citenamefont
  {Aspuru-Guzik},\ and\ \citenamefont {O'Brien}}]{Peruzzo_2014}%
  \BibitemOpen
  \bibfield  {author} {\bibinfo {author} {\bibfnamefont {A.}~\bibnamefont
  {Peruzzo}}, \bibinfo {author} {\bibfnamefont {J.}~\bibnamefont {McClean}},
  \bibinfo {author} {\bibfnamefont {P.}~\bibnamefont {Shadbolt}}, \bibinfo
  {author} {\bibfnamefont {M.-H.}\ \bibnamefont {Yung}}, \bibinfo {author}
  {\bibfnamefont {X.-Q.}\ \bibnamefont {Zhou}}, \bibinfo {author}
  {\bibfnamefont {P.~J.}\ \bibnamefont {Love}}, \bibinfo {author}
  {\bibfnamefont {A.}~\bibnamefont {Aspuru-Guzik}},\ and\ \bibinfo {author}
  {\bibfnamefont {J.~L.}\ \bibnamefont {O'Brien}},\ }\bibfield  {title}
  {\bibinfo {title} {A variational eigenvalue solver on a photonic quantum
  processor},\ }\href {https://doi.org/10.1038/ncomms5213} {\bibfield
  {journal} {\bibinfo  {journal} {Nature Communications}\ }\textbf {\bibinfo
  {volume} {5}},\ \bibinfo {pages} {4213} (\bibinfo {year} {2014})}\BibitemShut
  {NoStop}%
\bibitem [{\citenamefont {Tilly}\ \emph {et~al.}(2022)\citenamefont {Tilly},
  \citenamefont {Chen}, \citenamefont {Cao}, \citenamefont {Picozzi},
  \citenamefont {Setia}, \citenamefont {Li}, \citenamefont {Grant},
  \citenamefont {Wossnig}, \citenamefont {Rungger}, \citenamefont {Booth},\
  and\ \citenamefont {Tennyson}}]{TILLY20221}%
  \BibitemOpen
  \bibfield  {author} {\bibinfo {author} {\bibfnamefont {J.}~\bibnamefont
  {Tilly}}, \bibinfo {author} {\bibfnamefont {H.}~\bibnamefont {Chen}},
  \bibinfo {author} {\bibfnamefont {S.}~\bibnamefont {Cao}}, \bibinfo {author}
  {\bibfnamefont {D.}~\bibnamefont {Picozzi}}, \bibinfo {author} {\bibfnamefont
  {K.}~\bibnamefont {Setia}}, \bibinfo {author} {\bibfnamefont
  {Y.}~\bibnamefont {Li}}, \bibinfo {author} {\bibfnamefont {E.}~\bibnamefont
  {Grant}}, \bibinfo {author} {\bibfnamefont {L.}~\bibnamefont {Wossnig}},
  \bibinfo {author} {\bibfnamefont {I.}~\bibnamefont {Rungger}}, \bibinfo
  {author} {\bibfnamefont {G.~H.}\ \bibnamefont {Booth}},\ and\ \bibinfo
  {author} {\bibfnamefont {J.}~\bibnamefont {Tennyson}},\ }\bibfield  {title}
  {\bibinfo {title} {The variational quantum eigensolver: A review of methods
  and best practices},\ }\href
  {https://doi.org/https://doi.org/10.1016/j.physrep.2022.08.003} {\bibfield
  {journal} {\bibinfo  {journal} {Physics Reports}\ }\textbf {\bibinfo {volume}
  {986}},\ \bibinfo {pages} {1} (\bibinfo {year} {2022})},\ \bibinfo {note}
  {the Variational Quantum Eigensolver: a review of methods and best
  practices}\BibitemShut {NoStop}%
\bibitem [{\citenamefont {Akimoto}\ \emph {et~al.}(2023)\citenamefont
  {Akimoto}, \citenamefont {Kosuke}, \citenamefont {Leonardo}, \citenamefont
  {Takanori},\ and\ \citenamefont {Keisuke}}]{nakayama2023github}%
  \BibitemOpen
  \bibfield  {author} {\bibinfo {author} {\bibfnamefont {N.}~\bibnamefont
  {Akimoto}}, \bibinfo {author} {\bibfnamefont {M.}~\bibnamefont {Kosuke}},
  \bibinfo {author} {\bibfnamefont {P.}~\bibnamefont {Leonardo}}, \bibinfo
  {author} {\bibfnamefont {S.}~\bibnamefont {Takanori}},\ and\ \bibinfo
  {author} {\bibfnamefont {F.}~\bibnamefont {Keisuke}},\ }\href
  {https://github.com/Qulacs-Osaka/VQE-generated-dataset} {\bibinfo {title}
  {{VQE}-generated quantum circuit dataset}} (\bibinfo {year}
  {2023})\BibitemShut {NoStop}%
\bibitem [{\citenamefont {Cross}\ \emph {et~al.}(2022)\citenamefont {Cross},
  \citenamefont {Javadi-Abhari}, \citenamefont {Alexander}, \citenamefont
  {De~Beaudrap}, \citenamefont {Bishop}, \citenamefont {Heidel}, \citenamefont
  {Ryan}, \citenamefont {Sivarajah}, \citenamefont {Smolin}, \citenamefont
  {Gambetta},\ and\ \citenamefont {Johnson}}]{10.1145/3505636}%
  \BibitemOpen
  \bibfield  {author} {\bibinfo {author} {\bibfnamefont {A.}~\bibnamefont
  {Cross}}, \bibinfo {author} {\bibfnamefont {A.}~\bibnamefont
  {Javadi-Abhari}}, \bibinfo {author} {\bibfnamefont {T.}~\bibnamefont
  {Alexander}}, \bibinfo {author} {\bibfnamefont {N.}~\bibnamefont
  {De~Beaudrap}}, \bibinfo {author} {\bibfnamefont {L.~S.}\ \bibnamefont
  {Bishop}}, \bibinfo {author} {\bibfnamefont {S.}~\bibnamefont {Heidel}},
  \bibinfo {author} {\bibfnamefont {C.~A.}\ \bibnamefont {Ryan}}, \bibinfo
  {author} {\bibfnamefont {P.}~\bibnamefont {Sivarajah}}, \bibinfo {author}
  {\bibfnamefont {J.}~\bibnamefont {Smolin}}, \bibinfo {author} {\bibfnamefont
  {J.~M.}\ \bibnamefont {Gambetta}},\ and\ \bibinfo {author} {\bibfnamefont
  {B.~R.}\ \bibnamefont {Johnson}},\ }\bibfield  {title} {\bibinfo {title}
  {{OpenQASM 3: A Broader and Deeper Quantum Assembly Language}},\ }\bibfield
  {journal} {\bibinfo  {journal} {ACM Transactions on Quantum Computing}\
  }\textbf {\bibinfo {volume} {3}},\ \href {https://doi.org/10.1145/3505636}
  {10.1145/3505636} (\bibinfo {year} {2022})\BibitemShut {NoStop}%
\bibitem [{\citenamefont {Kandala}\ \emph {et~al.}(2017)\citenamefont
  {Kandala}, \citenamefont {Mezzacapo}, \citenamefont {Temme}, \citenamefont
  {Takita}, \citenamefont {Brink}, \citenamefont {Chow},\ and\ \citenamefont
  {Gambetta}}]{kandala2017hardware}%
  \BibitemOpen
  \bibfield  {author} {\bibinfo {author} {\bibfnamefont {A.}~\bibnamefont
  {Kandala}}, \bibinfo {author} {\bibfnamefont {A.}~\bibnamefont {Mezzacapo}},
  \bibinfo {author} {\bibfnamefont {K.}~\bibnamefont {Temme}}, \bibinfo
  {author} {\bibfnamefont {M.}~\bibnamefont {Takita}}, \bibinfo {author}
  {\bibfnamefont {M.}~\bibnamefont {Brink}}, \bibinfo {author} {\bibfnamefont
  {J.~M.}\ \bibnamefont {Chow}},\ and\ \bibinfo {author} {\bibfnamefont
  {J.~M.}\ \bibnamefont {Gambetta}},\ }\bibfield  {title} {\bibinfo {title}
  {Hardware-efficient variational quantum eigensolver for small molecules and
  quantum magnets},\ }\href@noop {} {\bibfield  {journal} {\bibinfo  {journal}
  {Nature}\ }\textbf {\bibinfo {volume} {549}},\ \bibinfo {pages} {242}
  (\bibinfo {year} {2017})}\BibitemShut {NoStop}%
\bibitem [{\citenamefont {Nakanishi}\ \emph {et~al.}(2019)\citenamefont
  {Nakanishi}, \citenamefont {Mitarai},\ and\ \citenamefont {Fujii}}]{ssvqe}%
  \BibitemOpen
  \bibfield  {author} {\bibinfo {author} {\bibfnamefont {K.~M.}\ \bibnamefont
  {Nakanishi}}, \bibinfo {author} {\bibfnamefont {K.}~\bibnamefont {Mitarai}},\
  and\ \bibinfo {author} {\bibfnamefont {K.}~\bibnamefont {Fujii}},\ }\bibfield
   {title} {\bibinfo {title} {Subspace-search variational quantum eigensolver
  for excited states},\ }\href
  {https://doi.org/10.1103/PhysRevResearch.1.033062} {\bibfield  {journal}
  {\bibinfo  {journal} {Phys. Rev. Res.}\ }\textbf {\bibinfo {volume} {1}},\
  \bibinfo {pages} {033062} (\bibinfo {year} {2019})}\BibitemShut {NoStop}%
\bibitem [{\citenamefont {K{\"{u}}bler}\ \emph {et~al.}(2020)\citenamefont
  {K{\"{u}}bler}, \citenamefont {Arrasmith}, \citenamefont {Cincio},\ and\
  \citenamefont {Coles}}]{Kubler2020adaptiveoptimizer}%
  \BibitemOpen
  \bibfield  {author} {\bibinfo {author} {\bibfnamefont {J.~M.}\ \bibnamefont
  {K{\"{u}}bler}}, \bibinfo {author} {\bibfnamefont {A.}~\bibnamefont
  {Arrasmith}}, \bibinfo {author} {\bibfnamefont {L.}~\bibnamefont {Cincio}},\
  and\ \bibinfo {author} {\bibfnamefont {P.~J.}\ \bibnamefont {Coles}},\
  }\bibfield  {title} {\bibinfo {title} {An {A}daptive {O}ptimizer for
  {M}easurement-{F}rugal {V}ariational {A}lgorithms},\ }\href
  {https://doi.org/10.22331/q-2020-05-11-263} {\bibfield  {journal} {\bibinfo
  {journal} {{Quantum}}\ }\textbf {\bibinfo {volume} {4}},\ \bibinfo {pages}
  {263} (\bibinfo {year} {2020})}\BibitemShut {NoStop}%
\bibitem [{\citenamefont {Nakanishi}\ \emph {et~al.}(2020)\citenamefont
  {Nakanishi}, \citenamefont {Fujii},\ and\ \citenamefont {Todo}}]{nft}%
  \BibitemOpen
  \bibfield  {author} {\bibinfo {author} {\bibfnamefont {K.~M.}\ \bibnamefont
  {Nakanishi}}, \bibinfo {author} {\bibfnamefont {K.}~\bibnamefont {Fujii}},\
  and\ \bibinfo {author} {\bibfnamefont {S.}~\bibnamefont {Todo}},\ }\bibfield
  {title} {\bibinfo {title} {Sequential minimal optimization for
  quantum-classical hybrid algorithms},\ }\href
  {https://doi.org/10.1103/PhysRevResearch.2.043158} {\bibfield  {journal}
  {\bibinfo  {journal} {Phys. Rev. Res.}\ }\textbf {\bibinfo {volume} {2}},\
  \bibinfo {pages} {043158} (\bibinfo {year} {2020})}\BibitemShut {NoStop}%
\bibitem [{\citenamefont {McClean}\ \emph {et~al.}(2018)\citenamefont
  {McClean}, \citenamefont {Boixo}, \citenamefont {Smelyanskiy}, \citenamefont
  {Babbush},\ and\ \citenamefont {Neven}}]{mcclean2018barren}%
  \BibitemOpen
  \bibfield  {author} {\bibinfo {author} {\bibfnamefont {J.~R.}\ \bibnamefont
  {McClean}}, \bibinfo {author} {\bibfnamefont {S.}~\bibnamefont {Boixo}},
  \bibinfo {author} {\bibfnamefont {V.~N.}\ \bibnamefont {Smelyanskiy}},
  \bibinfo {author} {\bibfnamefont {R.}~\bibnamefont {Babbush}},\ and\ \bibinfo
  {author} {\bibfnamefont {H.}~\bibnamefont {Neven}},\ }\bibfield  {title}
  {\bibinfo {title} {Barren plateaus in quantum neural network training
  landscapes},\ }\href@noop {} {\bibfield  {journal} {\bibinfo  {journal}
  {Nature communications}\ }\textbf {\bibinfo {volume} {9}},\ \bibinfo {pages}
  {4812} (\bibinfo {year} {2018})}\BibitemShut {NoStop}%
\bibitem [{\citenamefont {Parrish}\ \emph {et~al.}(2019)\citenamefont
  {Parrish}, \citenamefont {Hohenstein}, \citenamefont {McMahon},\ and\
  \citenamefont {Mart\'{\i}nez}}]{parrish2019}%
  \BibitemOpen
  \bibfield  {author} {\bibinfo {author} {\bibfnamefont {R.~M.}\ \bibnamefont
  {Parrish}}, \bibinfo {author} {\bibfnamefont {E.~G.}\ \bibnamefont
  {Hohenstein}}, \bibinfo {author} {\bibfnamefont {P.~L.}\ \bibnamefont
  {McMahon}},\ and\ \bibinfo {author} {\bibfnamefont {T.~J.}\ \bibnamefont
  {Mart\'{\i}nez}},\ }\bibfield  {title} {\bibinfo {title} {Quantum computation
  of electronic transitions using a variational quantum eigensolver},\ }\href
  {https://doi.org/10.1103/PhysRevLett.122.230401} {\bibfield  {journal}
  {\bibinfo  {journal} {Phys. Rev. Lett.}\ }\textbf {\bibinfo {volume} {122}},\
  \bibinfo {pages} {230401} (\bibinfo {year} {2019})}\BibitemShut {NoStop}%
\bibitem [{\citenamefont {Slattery}\ \emph {et~al.}(2022)\citenamefont
  {Slattery}, \citenamefont {Villalonga},\ and\ \citenamefont
  {Clark}}]{Slattery2022}%
  \BibitemOpen
  \bibfield  {author} {\bibinfo {author} {\bibfnamefont {L.}~\bibnamefont
  {Slattery}}, \bibinfo {author} {\bibfnamefont {B.}~\bibnamefont
  {Villalonga}},\ and\ \bibinfo {author} {\bibfnamefont {B.~K.}\ \bibnamefont
  {Clark}},\ }\bibfield  {title} {\bibinfo {title} {Unitary block optimization
  for variational quantum algorithms},\ }\href
  {https://doi.org/10.1103/PhysRevResearch.4.023072} {\bibfield  {journal}
  {\bibinfo  {journal} {Phys. Rev. Res.}\ }\textbf {\bibinfo {volume} {4}},\
  \bibinfo {pages} {023072} (\bibinfo {year} {2022})}\BibitemShut {NoStop}%
\bibitem [{\citenamefont {Dborin}\ \emph {et~al.}(2022)\citenamefont {Dborin},
  \citenamefont {Barratt}, \citenamefont {Wimalaweera}, \citenamefont
  {Wright},\ and\ \citenamefont {Green}}]{Dborin_2022}%
  \BibitemOpen
  \bibfield  {author} {\bibinfo {author} {\bibfnamefont {J.}~\bibnamefont
  {Dborin}}, \bibinfo {author} {\bibfnamefont {F.}~\bibnamefont {Barratt}},
  \bibinfo {author} {\bibfnamefont {V.}~\bibnamefont {Wimalaweera}}, \bibinfo
  {author} {\bibfnamefont {L.}~\bibnamefont {Wright}},\ and\ \bibinfo {author}
  {\bibfnamefont {A.~G.}\ \bibnamefont {Green}},\ }\bibfield  {title} {\bibinfo
  {title} {Matrix product state pre-training for quantum machine learning},\
  }\href {https://doi.org/10.1088/2058-9565/ac7073} {\bibfield  {journal}
  {\bibinfo  {journal} {Quantum Science and Technology}\ }\textbf {\bibinfo
  {volume} {7}},\ \bibinfo {pages} {035014} (\bibinfo {year}
  {2022})}\BibitemShut {NoStop}%
\bibitem [{\citenamefont {Rudolph}\ \emph {et~al.}(2023)\citenamefont
  {Rudolph}, \citenamefont {Miller}, \citenamefont {Motlagh}, \citenamefont
  {Chen}, \citenamefont {Acharya},\ and\ \citenamefont
  {Perdomo-Ortiz}}]{Rodolph2022}%
  \BibitemOpen
  \bibfield  {author} {\bibinfo {author} {\bibfnamefont {M.~S.}\ \bibnamefont
  {Rudolph}}, \bibinfo {author} {\bibfnamefont {J.}~\bibnamefont {Miller}},
  \bibinfo {author} {\bibfnamefont {D.}~\bibnamefont {Motlagh}}, \bibinfo
  {author} {\bibfnamefont {J.}~\bibnamefont {Chen}}, \bibinfo {author}
  {\bibfnamefont {A.}~\bibnamefont {Acharya}},\ and\ \bibinfo {author}
  {\bibfnamefont {A.}~\bibnamefont {Perdomo-Ortiz}},\ }\bibfield  {title}
  {\bibinfo {title} {Synergistic pretraining of parametrized quantum circuits
  via tensor networks},\ }\href@noop {} {\bibfield  {journal} {\bibinfo
  {journal} {Nature Communications}\ }\textbf {\bibinfo {volume} {14}},\
  \bibinfo {pages} {8367} (\bibinfo {year} {2023})}\BibitemShut {NoStop}%
\bibitem [{\citenamefont {Wecker}\ \emph {et~al.}(2015)\citenamefont {Wecker},
  \citenamefont {Hastings},\ and\ \citenamefont {Troyer}}]{Wecker2015}%
  \BibitemOpen
  \bibfield  {author} {\bibinfo {author} {\bibfnamefont {D.}~\bibnamefont
  {Wecker}}, \bibinfo {author} {\bibfnamefont {M.~B.}\ \bibnamefont
  {Hastings}},\ and\ \bibinfo {author} {\bibfnamefont {M.}~\bibnamefont
  {Troyer}},\ }\bibfield  {title} {\bibinfo {title} {Progress towards practical
  quantum variational algorithms},\ }\href
  {https://doi.org/10.1103/PhysRevA.92.042303} {\bibfield  {journal} {\bibinfo
  {journal} {Phys. Rev. A}\ }\textbf {\bibinfo {volume} {92}},\ \bibinfo
  {pages} {042303} (\bibinfo {year} {2015})}\BibitemShut {NoStop}%
\bibitem [{\citenamefont {Wiersema}\ \emph {et~al.}(2020)\citenamefont
  {Wiersema}, \citenamefont {Zhou}, \citenamefont {de~Sereville}, \citenamefont
  {Carrasquilla}, \citenamefont {Kim},\ and\ \citenamefont
  {Yuen}}]{Wierseme2020}%
  \BibitemOpen
  \bibfield  {author} {\bibinfo {author} {\bibfnamefont {R.}~\bibnamefont
  {Wiersema}}, \bibinfo {author} {\bibfnamefont {C.}~\bibnamefont {Zhou}},
  \bibinfo {author} {\bibfnamefont {Y.}~\bibnamefont {de~Sereville}}, \bibinfo
  {author} {\bibfnamefont {J.~F.}\ \bibnamefont {Carrasquilla}}, \bibinfo
  {author} {\bibfnamefont {Y.~B.}\ \bibnamefont {Kim}},\ and\ \bibinfo {author}
  {\bibfnamefont {H.}~\bibnamefont {Yuen}},\ }\bibfield  {title} {\bibinfo
  {title} {Exploring entanglement and optimization within the hamiltonian
  variational ansatz},\ }\href {https://doi.org/10.1103/PRXQuantum.1.020319}
  {\bibfield  {journal} {\bibinfo  {journal} {PRX Quantum}\ }\textbf {\bibinfo
  {volume} {1}},\ \bibinfo {pages} {020319} (\bibinfo {year}
  {2020})}\BibitemShut {NoStop}%
\bibitem [{\citenamefont {Virtanen}\ \emph {et~al.}(2020)\citenamefont
  {Virtanen}, \citenamefont {Gommers}, \citenamefont {Oliphant}, \citenamefont
  {Haberland}, \citenamefont {Reddy}, \citenamefont {Cournapeau}, \citenamefont
  {Burovski}, \citenamefont {Peterson}, \citenamefont {Weckesser},
  \citenamefont {Bright}, \citenamefont {{van der Walt}}, \citenamefont
  {Brett}, \citenamefont {Wilson}, \citenamefont {Millman}, \citenamefont
  {Mayorov}, \citenamefont {Nelson}, \citenamefont {Jones}, \citenamefont
  {Kern}, \citenamefont {Larson}, \citenamefont {Carey}, \citenamefont {Polat},
  \citenamefont {Feng}, \citenamefont {Moore}, \citenamefont {{VanderPlas}},
  \citenamefont {Laxalde}, \citenamefont {Perktold}, \citenamefont {Cimrman},
  \citenamefont {Henriksen}, \citenamefont {Quintero}, \citenamefont {Harris},
  \citenamefont {Archibald}, \citenamefont {Ribeiro}, \citenamefont
  {Pedregosa}, \citenamefont {{van Mulbregt}},\ and\ \citenamefont {{SciPy 1.0
  Contributors}}}]{2020SciPy-NMeth}%
  \BibitemOpen
  \bibfield  {author} {\bibinfo {author} {\bibfnamefont {P.}~\bibnamefont
  {Virtanen}}, \bibinfo {author} {\bibfnamefont {R.}~\bibnamefont {Gommers}},
  \bibinfo {author} {\bibfnamefont {T.~E.}\ \bibnamefont {Oliphant}}, \bibinfo
  {author} {\bibfnamefont {M.}~\bibnamefont {Haberland}}, \bibinfo {author}
  {\bibfnamefont {T.}~\bibnamefont {Reddy}}, \bibinfo {author} {\bibfnamefont
  {D.}~\bibnamefont {Cournapeau}}, \bibinfo {author} {\bibfnamefont
  {E.}~\bibnamefont {Burovski}}, \bibinfo {author} {\bibfnamefont
  {P.}~\bibnamefont {Peterson}}, \bibinfo {author} {\bibfnamefont
  {W.}~\bibnamefont {Weckesser}}, \bibinfo {author} {\bibfnamefont
  {J.}~\bibnamefont {Bright}}, \bibinfo {author} {\bibfnamefont {S.~J.}\
  \bibnamefont {{van der Walt}}}, \bibinfo {author} {\bibfnamefont
  {M.}~\bibnamefont {Brett}}, \bibinfo {author} {\bibfnamefont
  {J.}~\bibnamefont {Wilson}}, \bibinfo {author} {\bibfnamefont {K.~J.}\
  \bibnamefont {Millman}}, \bibinfo {author} {\bibfnamefont {N.}~\bibnamefont
  {Mayorov}}, \bibinfo {author} {\bibfnamefont {A.~R.~J.}\ \bibnamefont
  {Nelson}}, \bibinfo {author} {\bibfnamefont {E.}~\bibnamefont {Jones}},
  \bibinfo {author} {\bibfnamefont {R.}~\bibnamefont {Kern}}, \bibinfo {author}
  {\bibfnamefont {E.}~\bibnamefont {Larson}}, \bibinfo {author} {\bibfnamefont
  {C.~J.}\ \bibnamefont {Carey}}, \bibinfo {author} {\bibfnamefont
  {{\.I}.}~\bibnamefont {Polat}}, \bibinfo {author} {\bibfnamefont
  {Y.}~\bibnamefont {Feng}}, \bibinfo {author} {\bibfnamefont {E.~W.}\
  \bibnamefont {Moore}}, \bibinfo {author} {\bibfnamefont {J.}~\bibnamefont
  {{VanderPlas}}}, \bibinfo {author} {\bibfnamefont {D.}~\bibnamefont
  {Laxalde}}, \bibinfo {author} {\bibfnamefont {J.}~\bibnamefont {Perktold}},
  \bibinfo {author} {\bibfnamefont {R.}~\bibnamefont {Cimrman}}, \bibinfo
  {author} {\bibfnamefont {I.}~\bibnamefont {Henriksen}}, \bibinfo {author}
  {\bibfnamefont {E.~A.}\ \bibnamefont {Quintero}}, \bibinfo {author}
  {\bibfnamefont {C.~R.}\ \bibnamefont {Harris}}, \bibinfo {author}
  {\bibfnamefont {A.~M.}\ \bibnamefont {Archibald}}, \bibinfo {author}
  {\bibfnamefont {A.~H.}\ \bibnamefont {Ribeiro}}, \bibinfo {author}
  {\bibfnamefont {F.}~\bibnamefont {Pedregosa}}, \bibinfo {author}
  {\bibfnamefont {P.}~\bibnamefont {{van Mulbregt}}},\ and\ \bibinfo {author}
  {\bibnamefont {{SciPy 1.0 Contributors}}},\ }\bibfield  {title} {\bibinfo
  {title} {{{SciPy} 1.0: Fundamental Algorithms for Scientific Computing in
  Python}},\ }\href {https://doi.org/10.1038/s41592-019-0686-2} {\bibfield
  {journal} {\bibinfo  {journal} {Nature Methods}\ }\textbf {\bibinfo {volume}
  {17}},\ \bibinfo {pages} {261} (\bibinfo {year} {2020})}\BibitemShut
  {NoStop}%
\bibitem [{\citenamefont {Suzuki}\ \emph {et~al.}(2021)\citenamefont {Suzuki},
  \citenamefont {Kawase}, \citenamefont {Masumura}, \citenamefont {Hiraga},
  \citenamefont {Nakadai}, \citenamefont {Chen}, \citenamefont {Nakanishi},
  \citenamefont {Mitarai}, \citenamefont {Imai}, \citenamefont {Tamiya},
  \citenamefont {Yamamoto}, \citenamefont {Yan}, \citenamefont {Kawakubo},
  \citenamefont {Nakagawa}, \citenamefont {Ibe}, \citenamefont {Zhang},
  \citenamefont {Yamashita}, \citenamefont {Yoshimura}, \citenamefont
  {Hayashi},\ and\ \citenamefont {Fujii}}]{qulacs}%
  \BibitemOpen
  \bibfield  {author} {\bibinfo {author} {\bibfnamefont {Y.}~\bibnamefont
  {Suzuki}}, \bibinfo {author} {\bibfnamefont {Y.}~\bibnamefont {Kawase}},
  \bibinfo {author} {\bibfnamefont {Y.}~\bibnamefont {Masumura}}, \bibinfo
  {author} {\bibfnamefont {Y.}~\bibnamefont {Hiraga}}, \bibinfo {author}
  {\bibfnamefont {M.}~\bibnamefont {Nakadai}}, \bibinfo {author} {\bibfnamefont
  {J.}~\bibnamefont {Chen}}, \bibinfo {author} {\bibfnamefont {K.~M.}\
  \bibnamefont {Nakanishi}}, \bibinfo {author} {\bibfnamefont {K.}~\bibnamefont
  {Mitarai}}, \bibinfo {author} {\bibfnamefont {R.}~\bibnamefont {Imai}},
  \bibinfo {author} {\bibfnamefont {S.}~\bibnamefont {Tamiya}}, \bibinfo
  {author} {\bibfnamefont {T.}~\bibnamefont {Yamamoto}}, \bibinfo {author}
  {\bibfnamefont {T.}~\bibnamefont {Yan}}, \bibinfo {author} {\bibfnamefont
  {T.}~\bibnamefont {Kawakubo}}, \bibinfo {author} {\bibfnamefont {Y.~O.}\
  \bibnamefont {Nakagawa}}, \bibinfo {author} {\bibfnamefont {Y.}~\bibnamefont
  {Ibe}}, \bibinfo {author} {\bibfnamefont {Y.}~\bibnamefont {Zhang}}, \bibinfo
  {author} {\bibfnamefont {H.}~\bibnamefont {Yamashita}}, \bibinfo {author}
  {\bibfnamefont {H.}~\bibnamefont {Yoshimura}}, \bibinfo {author}
  {\bibfnamefont {A.}~\bibnamefont {Hayashi}},\ and\ \bibinfo {author}
  {\bibfnamefont {K.}~\bibnamefont {Fujii}},\ }\bibfield  {title} {\bibinfo
  {title} {Qulacs: a fast and versatile quantum circuit simulator for research
  purpose},\ }\href {https://doi.org/10.22331/q-2021-10-06-559} {\bibfield
  {journal} {\bibinfo  {journal} {{Quantum}}\ }\textbf {\bibinfo {volume}
  {5}},\ \bibinfo {pages} {559} (\bibinfo {year} {2021})}\BibitemShut {NoStop}%
\bibitem [{\citenamefont {van~der Maaten}\ and\ \citenamefont
  {Hinton}(2008)}]{vanDerMaaten2008}%
  \BibitemOpen
  \bibfield  {author} {\bibinfo {author} {\bibfnamefont {L.}~\bibnamefont
  {van~der Maaten}}\ and\ \bibinfo {author} {\bibfnamefont {G.}~\bibnamefont
  {Hinton}},\ }\bibfield  {title} {\bibinfo {title} {Visualizing data using
  t-sne},\ }\href@noop {} {\bibfield  {journal} {\bibinfo  {journal} {J. Mach.
  Learn. Res.}\ }\textbf {\bibinfo {volume} {9}},\ \bibinfo {pages} {2579}
  (\bibinfo {year} {2008})}\BibitemShut {NoStop}%
\bibitem [{\citenamefont {Kruskal}(1964)}]{Kruskal1964b}%
  \BibitemOpen
  \bibfield  {author} {\bibinfo {author} {\bibfnamefont {J.}~\bibnamefont
  {Kruskal}},\ }\bibfield  {title} {\bibinfo {title} {{Nonmetric
  multidimensional scaling: A numerical method}},\ }\href@noop {} {\bibfield
  {journal} {\bibinfo  {journal} {Psychometrika}\ }\textbf {\bibinfo {volume}
  {29}},\ \bibinfo {pages} {115–129} (\bibinfo {year} {1964})}\BibitemShut
  {NoStop}%
\bibitem [{\citenamefont {Novikov}(2019)}]{Novikov2019}%
  \BibitemOpen
  \bibfield  {author} {\bibinfo {author} {\bibfnamefont {A.}~\bibnamefont
  {Novikov}},\ }\bibfield  {title} {\bibinfo {title} {{PyClustering: Data
  Mining Library}},\ }\href {https://doi.org/10.21105/joss.01230} {\bibfield
  {journal} {\bibinfo  {journal} {Journal of Open Source Software}\ }\textbf
  {\bibinfo {volume} {4}},\ \bibinfo {pages} {1230} (\bibinfo {year}
  {2019})}\BibitemShut {NoStop}%
\bibitem [{\citenamefont {Hubert}\ and\ \citenamefont
  {Arabie}(1985)}]{hubert1985comparing}%
  \BibitemOpen
  \bibfield  {author} {\bibinfo {author} {\bibfnamefont {L.}~\bibnamefont
  {Hubert}}\ and\ \bibinfo {author} {\bibfnamefont {P.}~\bibnamefont
  {Arabie}},\ }\bibfield  {title} {\bibinfo {title} {Comparing partitions},\
  }\href@noop {} {\bibfield  {journal} {\bibinfo  {journal} {Journal of
  classification}\ }\textbf {\bibinfo {volume} {2}},\ \bibinfo {pages} {193}
  (\bibinfo {year} {1985})}\BibitemShut {NoStop}%
\bibitem [{\citenamefont {Wang}\ \emph {et~al.}(2022)\citenamefont {Wang},
  \citenamefont {Liu}, \citenamefont {Cheng}, \citenamefont {Liang},
  \citenamefont {Gu}, \citenamefont {Li}, \citenamefont {Ding}, \citenamefont
  {Jiang}, \citenamefont {Shi}, \citenamefont {Qian}, \citenamefont {Pan},
  \citenamefont {Chong},\ and\ \citenamefont {Han}}]{2210.16724v1}%
  \BibitemOpen
  \bibfield  {author} {\bibinfo {author} {\bibfnamefont {H.}~\bibnamefont
  {Wang}}, \bibinfo {author} {\bibfnamefont {P.}~\bibnamefont {Liu}}, \bibinfo
  {author} {\bibfnamefont {J.}~\bibnamefont {Cheng}}, \bibinfo {author}
  {\bibfnamefont {Z.}~\bibnamefont {Liang}}, \bibinfo {author} {\bibfnamefont
  {J.}~\bibnamefont {Gu}}, \bibinfo {author} {\bibfnamefont {Z.}~\bibnamefont
  {Li}}, \bibinfo {author} {\bibfnamefont {Y.}~\bibnamefont {Ding}}, \bibinfo
  {author} {\bibfnamefont {W.}~\bibnamefont {Jiang}}, \bibinfo {author}
  {\bibfnamefont {Y.}~\bibnamefont {Shi}}, \bibinfo {author} {\bibfnamefont
  {X.}~\bibnamefont {Qian}}, \bibinfo {author} {\bibfnamefont {D.~Z.}\
  \bibnamefont {Pan}}, \bibinfo {author} {\bibfnamefont {F.~T.}\ \bibnamefont
  {Chong}},\ and\ \bibinfo {author} {\bibfnamefont {S.}~\bibnamefont {Han}},\
  }\bibfield  {title} {\bibinfo {title} {Quest: Graph transformer for quantum
  circuit reliability estimation},\ }\Eprint
  {https://arxiv.org/abs/2210.16724v1} {arXiv:2210.16724v1 [quant-ph]}
  (\bibinfo {year} {2022})\BibitemShut {NoStop}%
\bibitem [{\citenamefont {Kipf}\ and\ \citenamefont
  {Welling}(2016)}]{kipf2016semisupervised}%
  \BibitemOpen
  \bibfield  {author} {\bibinfo {author} {\bibfnamefont {T.~N.}\ \bibnamefont
  {Kipf}}\ and\ \bibinfo {author} {\bibfnamefont {M.}~\bibnamefont {Welling}},\
  }\href@noop {} {\bibinfo {title} {Semi-supervised classification with graph
  convolutional networks}} (\bibinfo {year} {2016}),\ \Eprint
  {https://arxiv.org/abs/1609.02907} {arXiv:1609.02907 [cs.LG]} \BibitemShut
  {NoStop}%
\bibitem [{\citenamefont {Veličković}\ \emph {et~al.}(2018)\citenamefont
  {Veličković}, \citenamefont {Cucurull}, \citenamefont {Casanova},
  \citenamefont {Romero}, \citenamefont {Liò},\ and\ \citenamefont
  {Bengio}}]{veličković2018graph}%
  \BibitemOpen
  \bibfield  {author} {\bibinfo {author} {\bibfnamefont {P.}~\bibnamefont
  {Veličković}}, \bibinfo {author} {\bibfnamefont {G.}~\bibnamefont
  {Cucurull}}, \bibinfo {author} {\bibfnamefont {A.}~\bibnamefont {Casanova}},
  \bibinfo {author} {\bibfnamefont {A.}~\bibnamefont {Romero}}, \bibinfo
  {author} {\bibfnamefont {P.}~\bibnamefont {Liò}},\ and\ \bibinfo {author}
  {\bibfnamefont {Y.}~\bibnamefont {Bengio}},\ }\href@noop {} {\bibinfo {title}
  {Graph attention networks}} (\bibinfo {year} {2018}),\ \Eprint
  {https://arxiv.org/abs/1710.10903} {arXiv:1710.10903 [stat.ML]} \BibitemShut
  {NoStop}%
\bibitem [{\citenamefont {Xu}\ \emph {et~al.}(2019)\citenamefont {Xu},
  \citenamefont {Hu}, \citenamefont {Leskovec},\ and\ \citenamefont
  {Jegelka}}]{xu2019powerful}%
  \BibitemOpen
  \bibfield  {author} {\bibinfo {author} {\bibfnamefont {K.}~\bibnamefont
  {Xu}}, \bibinfo {author} {\bibfnamefont {W.}~\bibnamefont {Hu}}, \bibinfo
  {author} {\bibfnamefont {J.}~\bibnamefont {Leskovec}},\ and\ \bibinfo
  {author} {\bibfnamefont {S.}~\bibnamefont {Jegelka}},\ }\href@noop {}
  {\bibinfo {title} {How powerful are graph neural networks?}} (\bibinfo {year}
  {2019}),\ \Eprint {https://arxiv.org/abs/1810.00826} {arXiv:1810.00826
  [cs.LG]} \BibitemShut {NoStop}%
\bibitem [{\citenamefont {Hamilton}\ \emph {et~al.}(2017)\citenamefont
  {Hamilton}, \citenamefont {Ying},\ and\ \citenamefont
  {Leskovec}}]{hamilton2017inductive}%
  \BibitemOpen
  \bibfield  {author} {\bibinfo {author} {\bibfnamefont {W.~L.}\ \bibnamefont
  {Hamilton}}, \bibinfo {author} {\bibfnamefont {R.}~\bibnamefont {Ying}},\
  and\ \bibinfo {author} {\bibfnamefont {J.}~\bibnamefont {Leskovec}},\
  }\href@noop {} {\bibinfo {title} {Inductive representation learning on large
  graphs}} (\bibinfo {year} {2017}),\ \Eprint
  {https://arxiv.org/abs/1706.02216} {arXiv:1706.02216 [cs.SI]} \BibitemShut
  {NoStop}%
\bibitem [{\citenamefont {Platt}(2000)}]{PlattProbabilisticOutputs1999}%
  \BibitemOpen
  \bibfield  {author} {\bibinfo {author} {\bibfnamefont {J.}~\bibnamefont
  {Platt}},\ }\bibfield  {title} {\bibinfo {title} {Probabilistic outputs for
  support vector machines and comparison to regularized likelihood methods},\
  }in\ \href@noop {} {\emph {\bibinfo {booktitle} {Advances in Large Margin
  Classifiers}}}\ (\bibinfo {year} {2000})\BibitemShut {NoStop}%
\bibitem [{\citenamefont {Grant}\ \emph {et~al.}(2019)\citenamefont {Grant},
  \citenamefont {Wossnig}, \citenamefont {Ostaszewski},\ and\ \citenamefont
  {Benedetti}}]{grant2019initialization}%
  \BibitemOpen
  \bibfield  {author} {\bibinfo {author} {\bibfnamefont {E.}~\bibnamefont
  {Grant}}, \bibinfo {author} {\bibfnamefont {L.}~\bibnamefont {Wossnig}},
  \bibinfo {author} {\bibfnamefont {M.}~\bibnamefont {Ostaszewski}},\ and\
  \bibinfo {author} {\bibfnamefont {M.}~\bibnamefont {Benedetti}},\ }\bibfield
  {title} {\bibinfo {title} {An initialization strategy for addressing barren
  plateaus in parametrized quantum circuits},\ }\href@noop {} {\bibfield
  {journal} {\bibinfo  {journal} {Quantum}\ }\textbf {\bibinfo {volume} {3}},\
  \bibinfo {pages} {214} (\bibinfo {year} {2019})}\BibitemShut {NoStop}%
\bibitem [{\citenamefont {Skolik}\ \emph {et~al.}(2021)\citenamefont {Skolik},
  \citenamefont {McClean}, \citenamefont {Mohseni}, \citenamefont {van~der
  Smagt},\ and\ \citenamefont {Leib}}]{skolik2021layerwise}%
  \BibitemOpen
  \bibfield  {author} {\bibinfo {author} {\bibfnamefont {A.}~\bibnamefont
  {Skolik}}, \bibinfo {author} {\bibfnamefont {J.~R.}\ \bibnamefont {McClean}},
  \bibinfo {author} {\bibfnamefont {M.}~\bibnamefont {Mohseni}}, \bibinfo
  {author} {\bibfnamefont {P.}~\bibnamefont {van~der Smagt}},\ and\ \bibinfo
  {author} {\bibfnamefont {M.}~\bibnamefont {Leib}},\ }\bibfield  {title}
  {\bibinfo {title} {Layerwise learning for quantum neural networks},\
  }\href@noop {} {\bibfield  {journal} {\bibinfo  {journal} {Quantum Machine
  Intelligence}\ }\textbf {\bibinfo {volume} {3}},\ \bibinfo {pages} {1}
  (\bibinfo {year} {2021})}\BibitemShut {NoStop}%
\bibitem [{\citenamefont {Rudolph}\ \emph
  {et~al.}(2022{\natexlab{b}})\citenamefont {Rudolph}, \citenamefont {Miller},
  \citenamefont {Motlagh}, \citenamefont {Chen}, \citenamefont {Acharya},\ and\
  \citenamefont {Perdomo-Ortiz}}]{rudolph2022synergy}%
  \BibitemOpen
  \bibfield  {author} {\bibinfo {author} {\bibfnamefont {M.~S.}\ \bibnamefont
  {Rudolph}}, \bibinfo {author} {\bibfnamefont {J.}~\bibnamefont {Miller}},
  \bibinfo {author} {\bibfnamefont {D.}~\bibnamefont {Motlagh}}, \bibinfo
  {author} {\bibfnamefont {J.}~\bibnamefont {Chen}}, \bibinfo {author}
  {\bibfnamefont {A.}~\bibnamefont {Acharya}},\ and\ \bibinfo {author}
  {\bibfnamefont {A.}~\bibnamefont {Perdomo-Ortiz}},\ }\bibfield  {title}
  {\bibinfo {title} {Synergy between quantum circuits and tensor networks:
  Short-cutting the race to practical quantum advantage},\ }\href@noop {}
  {\bibfield  {journal} {\bibinfo  {journal} {arXiv preprint arXiv:2208.13673}\
  } (\bibinfo {year} {2022}{\natexlab{b}})}\BibitemShut {NoStop}%
\bibitem [{\citenamefont {Watanabe}\ \emph {et~al.}(2024)\citenamefont
  {Watanabe}, \citenamefont {Fujii},\ and\ \citenamefont
  {Ueda}}]{watanabe2023entangled}%
  \BibitemOpen
  \bibfield  {author} {\bibinfo {author} {\bibfnamefont {R.}~\bibnamefont
  {Watanabe}}, \bibinfo {author} {\bibfnamefont {K.}~\bibnamefont {Fujii}},\
  and\ \bibinfo {author} {\bibfnamefont {H.}~\bibnamefont {Ueda}},\ }\bibfield
  {title} {\bibinfo {title} {Variational quantum eigensolver with embedded
  entanglement using a tensor-network ansatz},\ }\href@noop {} {\bibfield
  {journal} {\bibinfo  {journal} {Physical Review Research}\ }\textbf {\bibinfo
  {volume} {6}},\ \bibinfo {pages} {023009} (\bibinfo {year}
  {2024})}\BibitemShut {NoStop}%
\end{thebibliography}%

\end{document}